\documentclass[journal]{IEEEtran}
\usepackage{bm, amssymb, graphicx, amsthm, color, caption, subcaption}
\usepackage[cmex10]{amsmath}

\usepackage{footnote}
\usepackage{perpage}
\makeatletter
\def\blfootnote{\xdef\@thefnmark{}\@footnotetext}
\makeatother

\newcommand{\bomega}{\bm{\omega}}

\newcommand{\bA}{\bm{A}}
\newcommand{\bB}{\bm{B}}

\newcommand{\bX}{\mathbf{X}}

\newcommand{\bD}{\mathbf{D}}
\newcommand{\bY}{\mathbf{Y}}
\newcommand{\bV}{\mathbf{V}}
\newcommand{\bZ}{\mathbf{Z}}

\newcommand{\bH}{\mathbf{H}}
\newcommand{\ba}{\mathbf{a}}
\newcommand{\by}{\mathbf{y}}
\newcommand{\bx}{\mathbf{x}}
\newcommand{\br}{\mathbf{r}}
\newcommand{\bu}{\mathbf{u}}
\newcommand{\bv}{\mathbf{v}}
\newcommand{\bz}{\mathbf{z}}

\newcommand{\bh}{\mathbf{h}}
\newcommand{\bb}{\mathbf{b}}

\newcommand{\eye}[1]{\mathbb{I}_{#1}}
\newcommand{\norm}[1]{\left\Vert{#1}\right\Vert}
\newcommand{\abs}[1]{\left|{#1}\right|}
\newcommand{\innerProd}[2]{\left<{#1}, {#2}\right>}
\newcommand{\fracc}[2]{\left.{#1}\middle/{#2}\right.}

\newcommand{\separation}{\vspace{1.5pt}}

\newcommand{\skeleton}[1]{}

\begin{document}
\title{Compressive channel estimation and tracking for large arrays in mm wave picocells} 

\author{Zhinus~Marzi,~
      Dinesh~Ramasamy~
        and~Upamanyu~Madhow,~\IEEEmembership{Fellow,~IEEE}\\
        \texttt{\{zh\_marzi, dineshr, madhow\}@ece.ucsb.edu}
\thanks{Z. Marzi, D. Ramasamy and U. Madhow are with the Department
of Electrical and Computer Engineering, University of California Santa Barbara, Santa Barbara,
CA}}

\maketitle

\begin{abstract}
We propose and investigate a compressive architecture for estimation and tracking of sparse spatial channels in millimeter (mm) wave picocellular networks. The base stations are equipped with antenna arrays with a large number of elements (which can fit within compact form factors because of the small carrier wavelength) and employ radio frequency (RF) beamforming, so that standard least squares adaptation techniques (which require access to individual antenna elements) are not applicable.  We focus on the downlink, and show that ``compressive beacons,'' transmitted using pseudorandom phase settings at the base station array, and compressively processed using pseudorandom phase settings at the mobile array,  provide information sufficient for accurate estimation of the two-dimensional (2D) spatial frequencies associated with the directions of departure of the dominant rays from the base station, and the associated complex gains. This compressive approach is  compatible with coarse phase-only control, and is based on a near-optimal sequential algorithm for frequency estimation which can exploit the geometric continuity of the channel across successive beaconing intervals to reduce the overhead  to less than 1\% even for very large ($32 \times 32$) arrays.  Compressive beaconing is essentially omnidirectional, and hence does not enjoy the SNR and spatial reuse benefits of beamforming obtained during data transmission. We therefore discuss system level design considerations for ensuring that the beacon SNR is sufficient for accurate channel estimation, and that inter-cell beacon interference is controlled by an appropriate reuse scheme.

\end{abstract}


\section{Introduction}
\label{sec:intro}

The explosive growth in demand for wireless mobile data, driven by the proliferation of ever more sophisticated handhelds creating and consuming rich multimedia, calls for orders of magnitude increases in the capacity of cellular data networks \cite{mobiledata}.  Millimeter wave communication from picocellular base stations to mobile devices is a particularly promising approach for meeting this challenge because of two reasons.  First, there are huge amounts of available spectrum, enabling channel bandwidths of the order of GHz, 1-2 orders of magnitude higher than those in existing systems at lower carrier frequencies. Indeed, channel bandwidths could potentially increase even further with advances in transceiver technology such as bandwidth/power consumption/linearity tradeoffs for ultra high-speed analog electronics, and speed/precision/power consumption tradeoffs for analog-to-digital converters. Second, the small carrier wavelength enables the realization of highly directive steerable arrays, with a large number of antenna elements, in compact form factors, thus significantly enhancing spatial reuse.  In this paper, we address fundamental signal processing challenges associated with channel estimation and tracking for such large arrays, placed within the context of system design for a mm wave picocellular network.

\begin{figure}
 \begin{center}
    \includegraphics[width=3.5in]{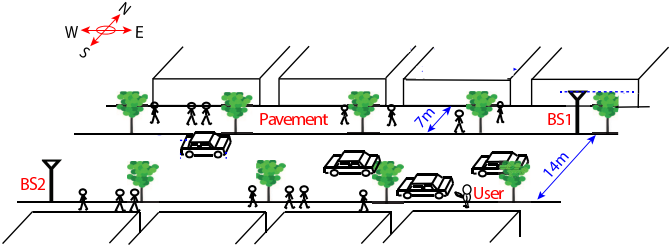}
\caption{Picocellular network deployed along an urban canyon}
\label{fig:urban}
\end{center}
\end{figure}

While the signal processing and system design concepts presented here are of rather general applicability, our numerical results are for a particular setting that we feel has great promise, as also discussed in some of our prior publications \cite{ramasamy_allerton12,Zhu_Mobicom2014}.  We propose to employ the 60 GHz unlicensed band for base station to mobile communication in outdoor picocells. The base stations can be deployed on lampposts, rooftops or ledges, and have multiple ``faces,'' with each face containing one or more antenna arrays.  An example deployment on lampposts in a zig-zag configuration (successive base stations on opposite sides of the street) along an urban canyon is shown in Figure \ref{fig:urban}. At the base station, we consider very large $32 \times 32$ arrays (such 1000-element arrays are still only palm-sized at a carrier wavelength of 5 mm) targeting the long term, as well as  ``moderately sized'' $8 \times 8$ arrays (which can fit within an area of about half a square inch) which are currently realizable.  Note that 16-element arrays were reported several years ago \cite{Valdes_2010}, and are already deployed in existing 60 GHz products, while 32-element arrays have been prototyped \cite{Cohen_2012}.  We assume that mobile devices are equipped with smaller $4 \times 4$ antenna arrays.  We focus on 60 GHz in order to leverage the significant advances that have occurred over the past few years targeting indoor wireless networks: once 60 GHz transceivers are embedded in mobile devices, one could use them to extend coverage to outdoor picocells, albeit with different approaches to medium access than in standard indoor wireless networking protocols. We focus on downlink 60 GHz communication, with the goal of enabling base station arrays to perform transmit beamforming towards mobile devices, despite the challenges posed by mobility and blockage (which occurs more easily at smaller wavelengths). We do not count on reciprocity. The uplink could be a standard LTE or WiFi link at lower carrier frequencies, used both for uplink data (not modeled here) and feedback for enabling spatial channel estimation at the transmitter.  

Multiple antenna systems at lower carrier frequencies have a relatively small number of elements, each with its own radio frequency (RF) chain. This provides control of the individual baseband signals associated with each element, enabling sophisticated adaptation, including frequency-selective spatiotemporal processing (e.g., per subcarrier beamforming in OFDM systems).  This approach does not scale when we have a large number of antenna elements packed into a tiny form factor. Instead, we consider RF beamforming, in which a common baseband signal is routed to/from the antenna elements, and we can only control the amplitude and phase for each element. Indeed, we go even further, assuming that the amplitude for each element is fixed, and that we can only apply coarse four-phase control for each element.  Standard least squares array adaptation and channel estimation techniques, which require access to the baseband signals associated with each element, do not apply in this setting.  Instead, we consider here a {\it compressive} approach which exploits the sparsity of the mm wave channel, so that relatively few measurements are required for channel estimation despite the large number of array elements.

\begin{figure}
\centering
\includegraphics[width=0.9\columnwidth]{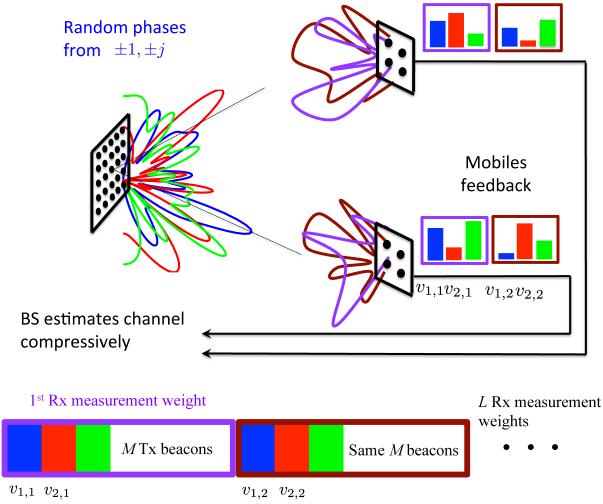}
\caption{Channel sounding scheme: The transmitter repeats the $M$ transmit beacons $L$ times so that receivers can measure the channel $v_{i,j}$ between every virtual transmit-receive pair}
\label{fig:channel_sounding}
\end{figure}

\noindent
{\bf Contributions:} Our contributions are summarized as follows:\\
{\it Architecture:} We propose a novel architecture in which base stations send out compressive beacons, with a different set of pseudorandom phases used to transmit each beacon. Each mobile measures the complex gains associated with each beacon compressively, using pseudorandom control of the phases of its receive array.  The scheme, described in more detail later, is depicted in Figure \ref{fig:channel_sounding}: the base station sends $M$ beacons, repeated $L$ times, which permits the mobile to use $L$ different settings of its own array to measure the associated complex gains. Each mobile feeds back appropriately chosen functions of its measurements to the base stations on the uplink (which can be a standard LTE link).  Each base station use this information to estimate and track the dominant paths to each mobile that it receives feedback from.  \\
{\it Algorithms:} For the regular two-dimensional (2D) arrays considered here, directions of arrival/departure map to 2D spatial frequencies. The base station estimates the spatial frequencies to each mobile using a simple sequential algorithm, shown to be near-optimal (in terms of approaching the Cramer-Rao Bound) in related publications.  The algorithm exploits the geometric continuity of the channel across successive beaconing intervals to reduce the required number of compressive measurements.\\
{\it System Design:} While we do not provide a complete system design centered around our compressive strategy, we do provide preliminary results addressing some of the most important issues. We show that the overhead incurred by our beaconing scheme is very small (less than 1\%). Furthermore, while compressive beacons are essentially omnidirectional (in contrast to the highly directive beams employed for communication), the link budget suffices for accurate channel estimation, and a simple beacon reuse strategy suffices to control inter-beacon interference across picocells.

\section{Background and Related Work} 

There is a significant body of recent literature on the potential for mm wave communication for next generation mobile cellular networks \cite{Rajagopal_Globecom_2011, Heath_asilomar2012, Rappaport_38G, Rappaport_WCNC2012, Rappaport_JSAC2014, Zhu_Mobicom2014}.  For these outdoor settings, most of the attention has focused on bands other than 60 GHz; for example, \cite{Rappaport_JSAC2014} studies blockage probabilities and achievable throughput based on measurements in the 28, 38, 71-76 and 81-86 GHz bands.  However, as pointed out in our prior work \cite{Zhu_Mobicom2014}, 60 GHz has immense potential at the relatively short ranges of interest in urban picocells, with the oxygen absorption characteristic of this band only having a marginal impact on the link budget (e.g., only 1.6 dB at 100m).  While studies such as \cite{Heath_asilomar2012, Rappaport_38G, Rappaport_WCNC2012, Rappaport_JSAC2014} focus on coarse-grained channel statistics (and their implications on system design and performance), our focus here is on signal processing for fine-grained channel estimation.  

We assume that the channel is well described by a relatively small number of discrete rays with delays and angles of arrival/departure taking values in a continuum.  The key contribution of this paper is to provide a super-resolution framework for estimating and tracking these rays, with model-based estimation allowing us to go beyond (spatial) bandwidth based resolution limits. To the best of our knowledge, the existing literature on mm wave measurements does not attempt to super-resolve channels in this fashion, hence we do not know, for example, whether the continuous power-delay profiles reported in \cite{Rappaport_JSAC2014} are consistent with a parsimonious channel model such as ours. However, preliminary experimental results \cite{Berraki_WCNC2014}, which use compressive measurements to successfully recover power-angle profiles for a controlled experiment (a small number of reflectors in an anechoic chamber), indicate that a simple ray-based model like ours may well suffice. Validating this assertion would require application of the more sophisticated compressive estimation techniques discussed here, as opposed to the standard basis pursuit algorithms employed in \cite{Berraki_WCNC2014}. Note that existing models for simulation-based evaluations, such as the channel models standardized for IEEE 802.11ad indoor 60 GHz channels, typically {\it assume} more complex models which are variants of the Saleh-Valenzuela model, with a number of clusters, each consisting of multiple closely spaced rays.  Such cluster-based models could be motivated by the roughness of reflecting surfaces such as walls, but they have not been experimentally demonstrated. While these issues fall beyond the scope of the present paper, which aims to make a fundamental contribution to signal processing for sparse spatial channels, a sustained effort in measurement-based validation of our model and approach is a critically important topic for future work.

Alternative approaches to spatial channel estimation with RF beamforming include codebook-based techniques; in an indoor WPAN setting, it was shown in \cite{Junyi_JSAC2009} that the minimum number of scanning beams for RF beamforming to get beamforming gains close to (within a dB) that of exhaustive search is twice the number of antenna elements. This approach does not scale to the large arrays of interest to us. Hierarchical codebook search \cite{david_love} is more efficient, but still requires far more overhead than our scheme, since it does not exploit channel sparsity.

A hardware enhancement to pure RF beamforming as considered here (where a single RF chain serves all antenna elements) is to employ hybrid analog-digital beamforming, with a number of RF chains smaller than the number of antenna elements.  We may term this an {\it array of subarrays,} with RF-level control for subarrays, and digital processing of subarray outputs. Such an approach is used in \cite{Heath_JSTSP2014,Heath_SPAWC_2014} for spatial channel estimation. Our work shows, however, that a single RF beamformed array suffices for this purpose.  Of course, arrays of subarrays are certainly required for more advanced functionalities such as multiuser MIMO \cite{Tadilo_Globecom2014}, spatial multiplexing \cite{torkildson,Gaojian2014,Singh_Globecom2014,Heath_ICC2012,Heath_SPAWC2012}, and spatial diversity \cite{Singh_Globecom2014,zhang2010statistical,zhang2010channel}. Integrating the compressive approach proposed in this paper within an array of subarrays architecture is an interesting area for future work.

The present paper builds on our prior conference papers on compressive array adaptation \cite{ramasamy_ita12,ramasamy_allerton12}, but goes well beyond them in several respects. In addition to a more detailed development of the analytical framework underlying our estimation algorithm, we now explicitly model the receive array at the mobile, which requires a generalization of the beaconing and feedback strategy.  We also address system level design for compressive tracking in far greater detail, discussing link budget and overhead, and accounting for inter-cell beacon interference.  Our initial work on compressive array adaptation \cite{ramasamy_ita12} subsequently led to a general framework for compressive estimation \cite{ramasamy_asilomar12,Ramasamy_TSP_2014}, which identify the isometries required to preserve fundamental bounds such as the Ziv-Zakai (ZZB) and Cramer-Rao (CRB), and use the relationship between these bounds to provide criteria for determining the minimum number of compressive measurements required to preserve geometry and to permit accurate parameter estimation based on a signal corrupted by an AWGN.  We now adapt these general results here to develop guidelines for system-level parameter choices.

\section{System Model}

Given the high demand for wireless data in dense urban environments, we focus our modeling and performance evaluation on the urban canyon setting depicted in Figure \ref{fig:urban}, with streets flanked by buildings on both sides. Picocellular base stations are deployed on lampposts in a zig-zag pattern on both sides of the street. We consider mm wave transmission on the downlink (for beaconing and downlink data) and LTE or WiFi at lower carrier frequencies on the uplink (for feedback and uplink data).  In terms of channel estimation and tracking, this could be viewed as a worst-case assumption, since two-way transmission on the same mm wave band could potentially be exploited using channel reciprocity. For the east-west street shown, each base station has two faces, facing east and west, respectively.  Each face can have multiple antenna arrays, but for simplicity, we consider a single antenna array for each face, used for both compressive beaconing and downlink data communication.  Mobile stations are modeled as either pedestrians walking along sidewalks, or cars moving along the street.

In our simulations, we model $K=4$ dominant paths from base station to mobile in our simulations: the line of sight (LoS), and the single bounce reflections from the ground or the side walls. Some of these paths may be blocked by obstacles (diffraction around obstacles is limited for small carrier wavelengths). We ignore multiple bounces, since such paths get attenuated significantly, especially because each bounce sees a smaller angle of incidence than a typical single bounce. However, our compressive estimation algorithm does not use the preceding assumptions on number of dominant paths as prior information, and automatically discovers and tracks paths.

For a regular $d$-spaced square $N_{1D} \times N_{1D}$ antenna array and a point transmitter in the far-field, the channel seen by the array is a mixture of 2D sinusoids, each corresponding to a propagation path, and is given by
\begin{equation*}
h_{m,n} = \sum_{l = 1}^{l = K} g_l e^{j \left(\omega_{x, l} m + \omega_{z,l} n\right)},~ g_l \in \mathbb{C},~ 1 \leq m,n, \leq N_{1D},
\end{equation*}
where $g_l$ is the propagation gain of the $l$-th path, $\bomega_l = \left(\omega_{x,l}, \omega_{z,l}\right)$ are the spatial frequencies corresponding to the $l$-th path (w.l.o.g. we have assumed that the square array is placed in the $x$-$z$ plane, with its sides aligned to the coordinate axes) and $h_{m,n}$ refers the channel seen by the $m,n$-th antenna element,. The spatial frequencies of the {$l$-th} path are given by $\omega_{x,l} = 2\pi (d/\lambda) \sin\theta_l\cos\phi_l$ and $\omega_{z,l} = 2\pi (d/\lambda) \sin\theta_l\sin\phi_l$, where $d$ denotes the array spacing, $\lambda$ the carrier wavelength and $\left(\theta_{l}, \phi_{l}\right)$ the inclination and azimuthal angles of the $l$-th path relative to $x-z$ plane. We vectorize the 2D sinusoid $\left[e^{j \left(\omega_{x} m + \omega_{z} n\right)},~0\leq m,n\leq N_{1D}-1\right]$ and denote the resulting $N_{1D}^2$ long vector by $\bx(N_{1D}, \bomega)$, where $\bomega = \left(\omega_{x}, \omega_{z}\right)$ is the frequency of the 2D sinusoid. Vectorizing $\left[h_{m,n},~1\leq m,n\leq N_{1D}\right]$ in an identical manner gives us
\begin{equation*}
\bh = \sum_{l = 1}^{l = K} g_l \bx(N_{1D}, \bomega_l).
\end{equation*}

Now, consider a base station transmitter with a regularly spaced 2D array of size $N_{t,1D}\times N_{t,1D}$, and a mobile receiver with a regular 2D antenna array of size $N_{r,1D}\times N_{r,1D}$. Let $\bH$ denote the corresponding $N_{t, 1D}^2 \times N_{r, 1D}^2$ channel matrix.  Denoting by $\bh_i$ the $i$th row of this matrix, $\bh_i^T$ is the response of the receive antenna array to the $i$th transmit antenna. Denoting $\bx\left(N_{t, 1D}, \bomega \right)$ by $\bx_t\left(\bomega \right)$ and $\bx\left(N_{r, 1D}, \bomega \right)$ by $\bx_r\left(\bomega \right)$, under the far-field assumption, it can be shown that
\begin{equation}\label{eq:channel_matrix}
\bH = \sum_{l = 1}^{l = K} g_l \bx_t\left(\bomega^{t}_{l}\right) \bx_r^T\left(\bomega^{r}_{l}\right),~ g_l\in\mathbb{C}.
\end{equation}
Since we know the array geometries (in this case, a regularly spaced 2D array), an estimate of the $N_{t, 1D}^2 \times N_{r, 1D}^2$ MIMO channel matrix $\bH$ can be efficiently arrived at by estimating the spatial frequencies and the associated gains: $\left\{\left(g_l, \bomega^{t}_{l}, \bomega^{r}_{l}\right),~l=1,\dots,K\right\}$. Such a parametric approach is far more efficient that direct estimation of individual entries of $\bH$, and enables us to drastically reduce the number of measurements required.

\section{Compressive channel estimation}\label{sec:channel_est}

We now describe our compressive approach for spatial channel estimation, which consists of a channel sounding strategy and an estimation algorithm which allows a base station to estimate the propagation gains  $\left\{ \left|g_l\right| \right\}$ and the spatial frequencies $\left\{ \bomega^{t}_{l} \right\}$ in parallel for \emph{all} mobiles in the picocell.

\subsection{Channel sounding} \label{sub-sec:channel_sounding}
The basestation sounds the channel using $M$ compressive beacons.  Each beacon is a known signal sent using a different set of transmit weights.  The weights are chosen uniformly and independently at random from a small set of coarse phase shifts (for e.g. from the set $\{\pm 1, \pm j\}$, where $j = \sqrt{-1}$).  We may therefore view each beacon as being transmitted from a different ``virtual antenna'' with a quasi-omnidirectional pattern. Each of the $M$ transmit beacons are repeated $L$ times by the basestation (see Figure~\ref{fig:channel_sounding}). For each of these $M$ transmit beacons, a mobile receiver employs $L$ ``virtual antennas'' to measure the channel response, using receive weights chosen uniformly at random from $\{\pm 1, \pm j\}$. Let $y(m,n)$ denote the response at the $(m,n)$th receive element due to a given transmit beacon. Letting $b({m,n,k}) \in \{\pm 1, \pm j\}$ denote the weight for $(m,n)$th receive element for the $k$th virtual receive antenna, the response seen by the $k$th virtual receive antenna is given by 
\begin{equation}\label{eq:receive_antenna-construction}
r(k) ~= \!\sum_{1\leq m, n\leq N_{r,1D}} \!\!b({m,n,k}) \times y({m,n}),~1\leq k\leq L.
\end{equation}
These measurements are used to construct the $M \times L$ Multiple Input Multiple Output (MIMO) ``virtual channel'' matrix $\bV$ between the virtual transmit and receive antennas. Note that we do not require that the base station know the receive weights used by the mobile, or that the mobile know the transmit weights used by the base station.

Denoting the vectorized version of weights of the $i$-th virtual transmit antenna by $\ba_i$ (a vector of shape $N_{t, 1D}^2 \times 1$) and that of the $j$-th virtual receive antenna by $\bb_j$ ($N_{r, 1D}^2 \times 1$), the $i,j$-th element of $\bV$ (the channel between the {$(i,j)$-th} virtual transmit-receive pair) is given by $v_{i,j} = \ba_i^T\bH\bb_j$. Letting $\bA = [\ba_1~\dots~\ba_{M}]^T$ and $\bB = [\bb_1~\dots~\bb_{L}]^T$, it is easy to see that $\bV = \bA \bH \bB^T$. Using \eqref{eq:channel_matrix}, we have that
\begin{equation}\label{eq:virtual_channel_matrix}
\bV =  \sum_{l = 1}^{l = K} g_l \left(\bA\bx_t\left(\bomega^{t}_{l}\right)\right) \left(\bB \bx_r\left(\bomega^{r}_{l}\right)\right)^T.
\end{equation}
The channel measurements are perturbed by i.i.d measurement noise, and are given by
\begin{equation*}
y_{i,j} = \sqrt{P_e} v_{i,j} + z_{i,j},~z_{i,j}\sim\mathcal{CN}(0, \sigma^2),
\end{equation*}
where ${P_e}$ is the per-element transmit power. Letting $\bY$ and $\bZ$ denote $M\times L$ matrices with their $i,j$-th entries given by $y_{i,j}$ and $z_{i,j}$ respectively, the ``measured virtual channel'' is given by
\begin{equation}\label{eq:virtual_channel_meas}
\bY  = \sqrt{P_e}\bV + \bZ = \sqrt{P_e}\sum_{l = 1}^{l = K} g_l \left(\bA\bx_t\left(\bomega^{t}_{l}\right)\right) \left(\bB \bx_r\left(\bomega^{r}_{l}\right)\right)^T + \bZ.
\end{equation}

\subsection{Feedback strategies} \label{sub-sec:channel_feedback}

Our goal is to track the mm-wave spatial channel as seen from the basestation, described by the parameters $P=\left\{\left(\left|g_l\right|, \bomega^t_l\right), l = 1,\dots, K\right\}$. To this end, every mobile in the picocell needs to feed back a portion of the measured virtual channel $\bY$ to the basestation. From \eqref{eq:virtual_channel_meas}, we see that the information regarding the spatial channel as seen from the basestation, given by $P$, is available in the \emph{column space} of $\bY$. Building on this observation, we propose two feedback strategies:\\
\noindent$~~$(i) The receiver feeds back the entire matrix $\bY$.\\ 
\noindent$~~$(ii) A more elaborate strategy involves feeding back $Q\leq L$ strongest left singular vectors of $\bY$ scaled by their corresponding singular values. i.e., if $\bY = \sum_{l=1}^{l=L}\sigma_l\bu_l\bv_l^H$ with $\sigma_1 \geq \sigma_2\geq\cdots\geq\sigma_L\geq 0$, the receivers feed back $\bD \equiv\left[\sigma_1\bu_1~\cdots~\sigma_Q\bu_Q\right]$. This strategy identifies the $Q$-dimensional \emph{subspace} of the column space of $\bY$ with maximum energy.


\section{Estimation Algorithm}\label{sec:algo}

We now present an algorithm to estimate the parameters $\left\{\left(\left|g_l\right|, \bomega^t_l\right), l = 1,\dots, K\right\}$ characterizing the mm-wave channel as seen from the basestation. The same algorithm applies for both forms of feedback discussed in Section~\ref{sub-sec:channel_feedback}: the entire measured virtual MIMO matrix $\bY$ or the dominant weighted left singular vectors $\bD$.

The $k$th column of $\bY$ is given by
\begin{equation}\label{eq:measurements_all}
\by_k = \sum_{l=1}^{l=K} h_{l, k} \bA\bx_t\left( \bomega^t_l \right) + \bz_k,~k = 1,\dots,L
\end{equation}
where $\bz_k\sim\mathcal{CN}\left(\mathbf{0}, \sigma^2\eye{M}\right)$ denotes the $k$-th column of $\bZ$ and $h_{l, k} = \sqrt{P_e}g_l \bb_k^T\bx_r\left(\bomega^r_l\right)$. We assume the weight sequence $\left\{\bb_k\right\}$ used to construct the receive virtual antennas at the receive antenna array is not available at the transmitter, and hence cannot jointly estimate $\bomega^t_l$ and $\bomega^r_l$. However, since $\{\bb_k,~k=1,\dots,L\}$ were picked in an i.i.d manner, we have that $\left\{h_{l, k},~k=1,\dots,L\right\}$ are i.i.d realizations of a random variable with $\mathbb{E}\left|h_{l, k}\right|^2= {P_e}\mathbb{E}\left|g_l\bb_k^T\bx_r\left(\bomega^r_l\right)\right|^2 = {P_e}N_{r,1D}^2\left|g_l\right|^2$. This allows us to estimate $\left|g_l\right|^2$ as follows:
$$
{P_e}\abs{\hat{g}_l}^2 = \left(\fracc{1}{\left(LN_{r,1D}^2\right)}\right)\sum_{k=1}^{k=L}\abs{\hat{h}_{l,k}}^2.
$$
From here on, in Section~\ref{sec:algo}, we use the notation $\bomega_l$ to refer to $\bomega^t_l$ and $\bx(\bomega_l)$ to refer to $\bA\bx_t(\bomega^t_l)$. Thus, the measurements can be written as
\begin{equation} \label{eq:measurements_simple}
\by_k = \sum_{l=1}^{l=K} h_{l, k} \bx\left( \bomega_l \right) + \bz_k,~k = 1,\dots,L.
\end{equation}
We now provide an algorithm to estimate $\{\bomega_l,\{h_{l,k}\}\}$. 

\subsection{Single path}\label{sub-sec:algo-single:path}
We first present an algorithm for estimating a single path $K=1$, which forms the basis for our sequential estimation algorithm for $K>1$. Omitting the path index $l$ in \eqref{eq:measurements_simple}, we have
\begin{equation*}
\by_k =  h_{k} \bx\left( \bomega \right) + \bz_k,~k = 1,\dots,L.
\end{equation*}
Given that $\{\bz_l\}$ are independent realizations of $\mathcal{CN}\left(\mathbf{0}, \sigma^2\eye{M}\right)$, the maximum likelihood (ML) estimates of $\bomega, \left\{h_{k}\right\}$ are given by:
\begin{equation}\label{eq:ML_est}
\hat{\bomega}, \left\{\hat{h}_{k}\right\} = \underset{\bomega, \left\{h_{k}\right\}}{\arg\min}\sum_{k = 1}^{k=L} \left\Vert \by_k - h_{k}\bx\left( \bomega \right) \right\Vert^2.
\end{equation}
For any $\bomega$, the optimal $h_{k}$-s are given by least-squares estimates:
\begin{equation}\label{eq:opt_gains}
h^{\star}_{k}\left(\bomega\right) = \fracc{\innerProd{\bx\left( {\bomega}\right)}{\by_k}}{\norm{\bx\left( {\bomega} \right)}^2}
\end{equation}
where $\langle x,y \rangle$ denotes $x^H y$. Plugging into \eqref{eq:ML_est}, the ML estimate of $\bomega$ is given by:
\begin{align}
\hat{\bomega} & = \underset{\bomega}{\arg\max}  ~~ \frac{1}{\left\Vert\bx\left( \bomega \right)\right\Vert^2}\sum_{k = 1}^{k=L} \left|\left<\bx\left( \bomega \right), \by_k \right>\right|^2\label{eq:GLRT_omega}
\end{align}
and the ML estimate of $h_{k}$ is given by $h^{\star}_{k}\left(\hat{\bomega}\right)$. We employ a two-step algorithm to arrive the ML estimates: a ``detection'' phase followed by a ``refinement'' phase.
\separation\\
\noindent\textbf{Detection phase:} Using $M$ two-dimensional FFT computations, we precompute $\bx(\bomega)$ for frequencies of the form $\bomega \in \Phi \equiv \left\{ \left(2\pi k/T, 2\pi l / T\right),~0\leq k,l\leq T-1\right\},~T = RN_{1D,t}$, where $R$ is the oversampling factor. We pick the frequency $\hat{\bomega}$ from $\Phi$ which maximizes \eqref{eq:GLRT_omega}. The corresponding gains are given by $\hat{h}_{k} = {h}^{\star}_{k}\left(\hat{\bomega}\right)$. We remove the contribution of the newly detected sinusoid from the measured channel response and this residual measurement is given by 
\begin{equation}\label{eq:residue_single}
\br_{k} = \by_k - \hat{h}_{k} \bx\left(\hat{\bomega}\right).
\end{equation}
(This residue is used for sequential detection for $K>1$, as discussed shortly.)

\separation
\noindent\textbf{Refinement phase:} Our estimate from the detection phase is restricted to the discrete set $\Phi$ and consequently we do not expect $\hat{\bomega}$ to be equal to the ML estimate given by \eqref{eq:GLRT_omega} (where the maximization is over $[-2\pi d/\lambda,2\pi d/\lambda]^2$ with $d$ being the  spacing between transmitter antennas). However, if we make the grid fine enough, the best estimate of $\bomega$ in $\Phi$ is expected to be close enough to the optimal solution to allow refinement via local optimization. In order to do this, we first \emph{fix} the gain estimates $\{\hat{h}_{k}\}$ and refine only the estimate of the spatial frequency $\bomega$ by seeking the minimizer of the ML cost function
\begin{equation*}
C(\bomega) = \sum_{k = 1}^{k=L} \left\Vert \by_k - \hat{h}_{k}\bx\left( \bomega \right) \right\Vert^2
\end{equation*}
in the neighborhood of the current estimate $\hat{\bomega}$ using the Newton method. This involves evaluating the gradient vector $\mathcal{G}\left(\bomega\right)$ and the Hessian matrix $\mathcal{H}(\bomega)$ of $C(\bomega)$ at the current estimate $\hat{\bomega}$. The corresponding expressions are given by:
\begin{align*}
\mathcal{G}_i(\bomega) & =\frac{\partial C(\bomega)}{\partial\omega_i} = -  2 \sum_{k = 1}^{k=L} \Re\left\{ \innerProd{ \br_k }{ \hat{h}_{k}\frac{\partial \bx(\bomega)}{\partial\omega_i} }\right\},\\
\mathcal{H}_{ij}(\bomega) & =\frac{\partial^2 C(\bomega)}{\partial\omega_i\partial\omega_j} = -  2 \sum_{k = 1}^{k=L} \Re\Bigg\{ \innerProd{ \br_k }{ \hat{h}_{k}\frac{\partial^2 \bx(\bomega)}{\partial\omega_i\partial\omega_j}} - \\
&~~~~~~~  \abs{\hat{h}_{k}}^2 \innerProd{ \frac{\partial \bx(\bomega)}{\partial\omega_i} }{ \frac{\partial \bx(\bomega)}{\partial\omega_j} }\Bigg\},~1\leq i,j\leq 2
\end{align*}
where $\bomega = [\omega_1,\omega_2]$. $\{\br_k\}$-s are the residual measurements given by \eqref{eq:residue_single}. The Newton update for $\hat{\bomega}$ is
\begin{equation}\label{eq:refine_omega}
\hat{\bomega} \gets \hat{\bomega} - \mathcal{H}^{-1} \left(\hat{\bomega}\right) \mathcal{G}\left(\hat{\bomega}\right).
\end{equation}
We follow this up by updating our estimates $\left\{\hat{h}_{k}\right\}$ by plugging the new value of $\hat{\bomega}$ in \eqref{eq:opt_gains}, i.e., 
\begin{equation}\label{eq:update_gain}
\hat{h}_{k} \gets h^{\star}_k\left(\hat{\bomega}\right) = \fracc{ \innerProd{\bx\left(\hat{\bomega} \right) }{ \by_k } }{ \norm{ \bx\left( \hat{\bomega} \right) }^2 }
\end{equation}
and modifying the residues ($\{\br_k\}$) accordingly using \eqref{eq:residue_single}. The algorithm alternates between the updates in \eqref{eq:refine_omega} and \eqref{eq:update_gain} for a few iterations. 

\subsection{Multiple paths}\label{sub-sec:algo-multiple:paths}
We now build on the preceding single path algorithm for the general setting of $K \geq 1$. Suppose that our current estimate of the sinusoids/paths is given by $\mathcal{P}_{K} = \left\{\left(\hat{\bomega}_l, \{\hat{h}_{l, k}\}\right),~l=1,\dots,K\right\}$. The residual measurements corresponding to a set of estimated parameters $\mathcal{P}$ is given by: 
$$ \textstyle
\bv_k(\mathcal{P}) = \by_k - \sum_{{\bomega}_l, \{{h}_{l, k}\} \in \mathcal{P}} {h}_{l, k} \bx\left({\bomega}_l\right).
$$

\noindent\textbf{Detect a new path:} Assuming that the measurements $\by_k$ are given by the current residue $\bv_k(\mathcal{P}_{q})$ (corresponding to the $q$ detected paths), we use the single path algorithm in Section~\ref{sub-sec:algo-single:path} to detect and refine a new sinusoid $(\hat{\bomega}_{q+1}, \{\hat{h}_{q+1, k}\})$. Let $\mathcal{P}_{q+1}$ denote the new set of estimated parameters $\mathcal{P}_q \cup \{(\hat{\bomega}_{q+1}, \{\hat{h}_{q+1, k}\})\}$. 

\separation
\noindent\textbf{Refine existing paths:} Once we add this new path, we refine the parameters of \emph{all} $q+1$ sinusoids in $\mathcal{P}_{q+1}$ one by one. Consider the parameters $(\hat{\bomega}_{l}, \{\hat{h}_{l, k}\})$ of the $l$-th sinusoid. We use the refinement algorithm in Section~\ref{sub-sec:algo-single:path} to refine $(\hat{\bomega}_{l}, \{\hat{h}_{l, k}\})$ by assuming that the measurements $\by_k$ are given by the residual measurements after excluding the sinusoid of interest. i.e, $\bv_k(\mathcal{P}_{q+1}\setminus \{(\hat{\bomega}_{l}, \{\hat{h}_{l, k}\})\})$. Sinusoids are refined in a round robin manner, and the process is repeated for a few rounds: $1\rightarrow 2\rightarrow\dots\rightarrow(q+1)\rightarrow 1\rightarrow\dots\rightarrow(q+1)$.

\separation
\noindent\textbf{Stopping criterion:} The algorithm continues to add newly detected paths as long as the payoff, determined by the total amount by which the residue decreases is above a threshold $\tau$. i.e., if
\begin{equation}\label{eq:stopping_criterion}
\sum_{k = 1}^{k = L} \left( \norm{ \bv_k(\mathcal{P}_{q}) }^2 - \norm{ \bv_k(\mathcal{P}_{q+1}) }^2 \right) > \tau,
\end{equation}
the $(q+1)$-th path is added and the algorithm proceeds by searching for a new path. On the contrary, if the reduction in total residue is smaller than $\tau$, the algorithm terminates and returns the prior estimate of the parameters $\mathcal{P}_{q}$. For the stopping criterion, we use $\tau = 30 \sigma^2 \log\left(20N_{t,1D}\right)$ in our simulations. This stopping criterion is empirically determined, such that the signal energy $\tau$ is large enough to comfortably cross the ZZB threshold (so that we can expect the frequency estimate to be accurate).

\subsection{Tracking}\label{sub-sec:algo-tracking}
We sound the channel often enough so that between any two successive channel estimation cycles, the \emph{geometry} of the mm-wave channel, given by the spatial frequencies $\{\bomega_l\}$ of the paths, do not change ``significantly,'' even if the path gains $\{g_l\}$ do. This ensures that angle of departure estimates from the prior sounding round do not become stale over the course of the communication phase during which they are needed for beamforming purposes. For example, if we do not wish to tolerate a beamforming loss of $3$dB or more, then our estimate from the previous round $\hat{\bomega}$ should be close enough to the current $\bomega$ so that 
$$
\fracc{\abs{\innerProd{\bx\left(\hat{\bomega}\right)}{\bx\left({\bomega}\right)}}^2}{\norm{\bx\left({\bomega}\right)}^2} > 0.5
$$
over the entire communication phase. This condition is met if $\norm{{\bomega}-\hat{\bomega}}_\infty < 0.5 \times \left(\fracc{2\pi}{N_{t, 1D}}\right)$. Therefore, the estimates of spatial frequencies from the previous sounding round are good approximations of their current true value (within a DFT spacing of $\frac{2\pi}{N_{t, 1D}}$). We exploit this by using $\{\hat{\bomega}_l,~l=1,\dots, K\}$ from the prior round to initialize our algorithm (as opposed to using the empty set $\{\}$). We do this by constructing the matrix $\bX = \left[ \bx\left(\hat{\bomega}_1\right)~\dots~\bx\left(\hat{\bomega}_K\right) \right]$ and setting $\hat{h}_{i,j}$ to be the $(i,j)$-th entries of $\left(\bX^H\bX\right)^{-1}\bX^H\bY$, where $\bY = [\by_1~\dots~\by_L]$. We refine all parameters in $\mathcal{P}_K = \{(\hat{\bomega}_{i}, \{\hat{h}_{i, j}\}),~i=1,\dots, K\}$ using the refinement algorithm in Section~\ref{sub-sec:algo-multiple:paths} before proceeding to seek for new paths using the algorithm in Section~\ref{sub-sec:algo-multiple:paths}.

\noindent\textit{Deleting weak paths:} Paths estimated in prior rounds may not be viable at the current time instant (e.g, because of blockage). Therefore, we need means to remove such stale paths. We use the stopping criterion \eqref{eq:stopping_criterion} as a means to delete weak paths. If deleting the path under question and optimizing other parameters increases the residue by an amount smaller than $\tau$, we delete the path permanently. Otherwise we keep the path.

\section{Protocol Parameter Choices}\label{sec:protocol_choice}

In this section, we give a principled approach to choosing parameters of the compressive channel estimation protocol, namely the number of unique transmit beacons $M$, the number of receive measurement weights $L$ and the minimum effective Signal to Noise Ratio (SNR) needed for channel estimation, which we use to choose the sounding bandwidth $W_s$. We then turn to the question of how frequently the channel has to be sounded. In Section~\ref{sec:sims}, we take two scenarios and apply this recipe to arrive at system level parameters for the protocol.

\subsection{Number of compressive transmit beacons}\label{sub-sec:geometry_M}

Our goal is to estimate the spatial frequencies $\{\bomega^t_l\}$ from measurements of the form 
$$
\by_k = \sum_{l=1}^{l=K} h_{l, k} \bA\bx_t\left( \bomega^t_l \right) + \bz_k,~k = 1,\dots,L.
$$ 
The algorithm in Section~\ref{sec:algo} aims to estimate parameters $\left\{{h}_{i,j}, {\bomega}_i\right\}$ by minimizing the ML cost function:
\begin{align*}
&\textstyle\!\!\!\sum_{j=1}^{j=L} \norm{\by_j - \sum_{i=1}^{i={K}} \hat{h}_{i,j}\bA\bx_t\left(\hat{\bomega}_i\right)}^2\\
&\textstyle= \sum_{j=1}^{j=L} \norm{\bA \times \sum_{i=1}^{i={K}} \left({h}_{i,j}\bx_t\left({\bomega}^t_i\right) -  \hat{h}_{i,j}\bx_t\left(\hat{\bomega}^t_i\right)\right) + \bz_j}^2,
\end{align*}
where $\hat{p}$ denotes an estimate of parameter $p$. If the compressive measurement matrix $\bA$ ensures that
\begin{align}
&\textstyle\!\!\!\norm{\bA\times \sum_{i=1}^{i={K}} \left( {h}_{i,j}  \bx_t\left({\bomega}^t_i\right) -  \hat{h}_{i,j}\bx_t\left(\hat{\bomega}^t_i\right)  \right)}^2\label{eq:geometry_preservation}\\
&\textstyle\quad \approx M \norm{\sum_{i=1}^{i={K}}\left( {h}_{i,j}\bx_t\left({\bomega}^t_i\right) - \hat{h}_{i,j}\bx_t\left(\hat{\bomega}^t_i \right) \right)}^2,~\forall {h}_{i,j}, \hat{h}_{i,j}, \nonumber
\end{align}
for relevant $(\{{\bomega}_i\}, \{\hat{\bomega}_i\})$-pairs, the cost structure of the estimation problem is roughly preserved. Therefore, estimation using compressive measurements is similar to estimation with all $N_{t,1D}^2$ measurements (except for an SNR gain, given by $M$) \cite{Ramasamy_TSP_2014}. The main idea behind compressive sensing is that a ``small'' number of random projections can ensure that the \emph{geometry preservation} condition in \eqref{eq:geometry_preservation} is met with high probability. Such results build on the celebrated Johnson-Lindenstrauss Lemma (JL Lemma).

\separation
\noindent\textit{JL lemma}\cite{achlioptas_database-friendly_2001}:~Consider a \emph{finite} collection of points $S \subset \mathbb{R}^n$ and an $m\times n$ matrix $\Phi$ with its entries chosen in an i.i.d. manner from $\textrm{Uniform}\{\pm 1\}$ with $m \geq \fracc{(4+2\beta)\log |S|}{\left(\epsilon^2/2 - \epsilon^3/3\right)}$, then with probability at least $1-|S|^{-\beta}$:
\begin{equation}\label{eq:JL:lemma}
m(1-\epsilon) \leq \frac{\norm{\Phi\bu - \Phi\bv}^2}{\norm{\bu - \bv}^2} \leq m(1+\epsilon)~\forall \bu,\bv\in S.
\end{equation}
It can be shown that this result extends to the setting where $S \subset \mathbb{C}^{n}$ and the entries of $\Phi$ are chosen from $\textrm{Uniform}\{\pm 1,\pm j\}$.

\begin{figure}
\centering
\includegraphics[width=0.7\columnwidth]{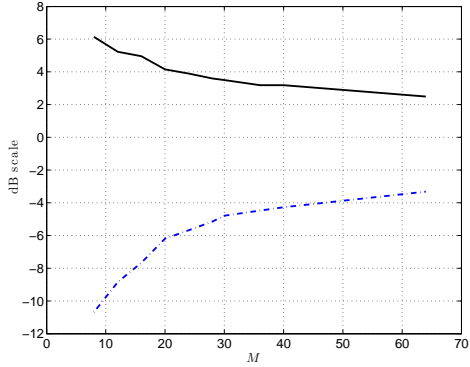}
\caption{Maximum and minimum values of $\fracc{\norm{\bA\bX\bu}^2}{(M\norm{\bX\bu}^2)}$ for different values of $M$, the number of transmitter beacons, across $5\times10^6$ random realizations of $8$-sparse $\bu$. The basis $\bX$ corresponds to the responses for a $32 \times 32$ array evaluated uniformly over a $R = 64\times 32^2$-sized grid}
\label{fig:choice_of_M}
\end{figure}

Returning to the geometry preservation condition in \eqref{eq:geometry_preservation}, we see that when spatial frequencies are restricted to an oversampled DFT grid $G$ of size $R = O(N_{t,1D}^2)$, this condition reduces to a $2K$-\emph{isometry property} of the measurement matrix $\bA$ relative to the basis $\bX$, where $\bX$ is the $N_{t,1D}^2 \times R$ matrix with its columns given by $\{\bx_t\left(\bomega\right) : \bomega\in G\}$. For some fixed $\epsilon$, the matrix $\Phi\in\mathbb{C}^{m\times p}$ is said to enjoy the $s$-isometry property for the basis $B$ (of size $p \times n$) if there exists a constant $C > 0$ such that
$$
C(1-\epsilon) \leq \fracc{\norm{\Phi B \bu}^2}{\norm{B \bu}^2} \leq C(1+\epsilon),
$$
for all $s$-sparse $\bu$ in $\mathbb{C}^n$. It can be shown using the JL lemma (with arguments similar to those in \cite{baraniuk_simple_RIP_2008}) that if $m = O(s\epsilon^{-2} \log n)$, a randomly picked $\Phi$ satisfies this isometry property with high probability (w.h.p). Therefore, when the number of unique transmitter beacons scales as $M = O\left(K\epsilon^{-2}\log R \right) = O\left(K\epsilon^{-2}\log N_{t,1D} \right)$, then the $2K$-pairwise isometry criterion w.r.t the basis $\bX$ is met by the randomly picked sounding matrix $\bA$ (it can be shown that $C = M$ for our choice of scale), thereby ensuring that the geometry of the spatial frequency estimation problem is preserved.

We consider the example of the $32 \times 32 $ transmitter array and plot the maximum and minimum values of $(1/M)\fracc{\norm{\bA \bX \bu}^2}{\norm{\bX \bu}^2}$ from $5\times 10^6$ random realizations of a $2K=8$-sparse $\bu$ in Figure~\ref{fig:choice_of_M} using the $64$-times oversampled DFT grid as the choice of basis $\bX$. We see that this ratio is within $[-5,3]$ dB when $M \geq 30$. This illustrates that for estimating $K = 4$ paths using a $32 \times 32$ array, measuring the response corresponding to $M=30$ random beacons approximates the effect of measuring all $32 \times 32 = 1024$ antenna elements individually.

\subsection{Number of compressive receive measurements}\label{sub-sec:geometry_L}

\begin{figure}
\centering
\includegraphics[width=0.7\columnwidth]{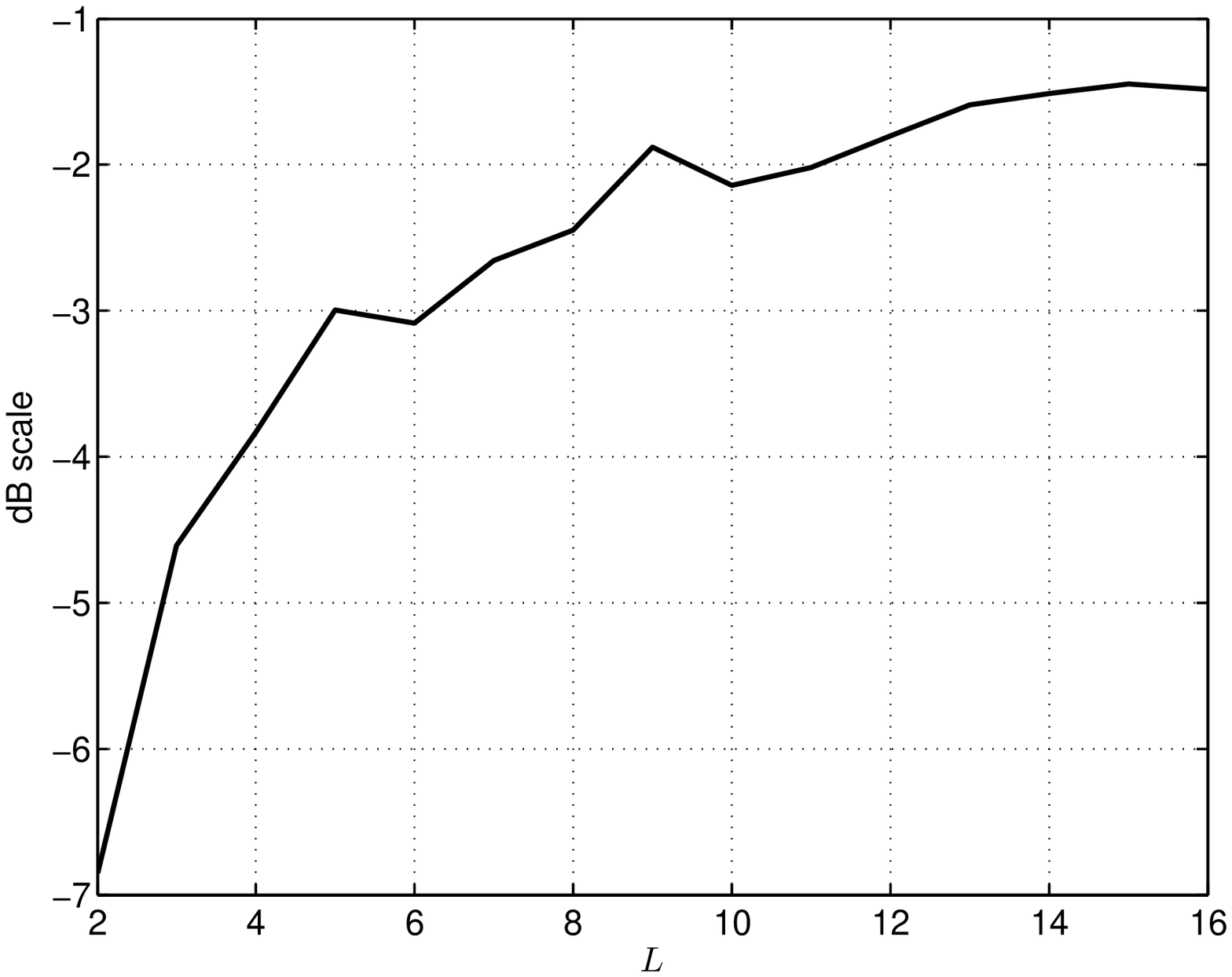}
\caption{Maximum SNR degradation $\min_{\bomega} \fracc{\norm{\bB\bx_r(\bomega)}^2}{(L\norm{\bx_r(\bomega)}^2)}$ for the most favorable realization (from $10^4$ runs) of an $L \times N_{r,1D}^2$ matrix $\bB$ with $N_{r,1D} = 4$}
\label{fig:choice_of_L}
\end{figure}

While we do not track $\{\bomega^r_i,~i=1,\dots,K\}$, the spatial frequencies at the receiver, we need to ensure that the set of measurements made at the receiver have sufficient information to estimate transmitter spatial frequencies. Suppose that $\norm{\bB\bx_r\left(\bomega^r_i\right)}\approx 0$, it follows from $h_{i, j} = g_i\sqrt{P_e} \bb_j^T\bx_r\left(\bomega^r_i\right)$ that all $L$ measurements $\{\by_j,1\leq j\leq L\}$ will have very small contributions from the $i$-th path. i.e., $\abs{h_{i,j}}\approx 0,1\leq j\leq L$. To see this observe that 
\begin{equation}\label{eq:relate_h_B}
\sum_{j=1}^{j=L}\abs{h_{i,j}}^2 = {P_e}\abs{g_i}^2\sum_{j=1}^{j=L} \abs{\bb_j^T\bx_r\left(\bomega^r_i\right)}^2 = {P_e}\abs{g_i}^2\norm{\bB\bx_r\left(\bomega^r_i\right)}^2.
\end{equation}
When we restrict the receive spatial frequencies to an oversampled DFT grid $G$ of size $R = O(N_{r,1D}^2)$, it can be shown that for $L = O\left( \log R\right) = O\left( \log N_{r, 1D} \right)$, $\norm{\bB \bx_r\left(\bomega\right)}^2 \approx L \norm{ \bx_r\left(\bomega\right)}^2 = L N_{r,1D}^2$ w.h.p (a direct application of the JL lemma \eqref{eq:JL:lemma} with the set $S$ being $\{\bx_r\left(\bomega\right):\bomega\in G\}\cup\{\mathbf{0}\}$, where $\mathbf{0}$ denotes the zero-vector). This ensures that 
\begin{equation*}
\sum_{j=1}^{j=L}\abs{h_{i,j}}^2 \approx {P_e}LN_{r,1D}^2\abs{g_i}^2~\text{w.h.p}, 
\end{equation*}
thereby capturing power along the $i$-th path. We perform computations for the maximum power lost across spatial frequencies when using a $4\times 4$ array and plot the results in Figure~\ref{fig:choice_of_L}. This shows that around $5$ carefully chosen projections (we pick the best measurement matrix from $10^4$ random instances) suffice to ensure that SNR degradation (relative to the nominal value of $L$) is no greater than $3$dB for a $4\times 4$ receive array.

\subsection{SNR for successful estimation}\label{sub-sec:SNR_needed}

\begin{figure}
\centering
\includegraphics[width=0.7\columnwidth]{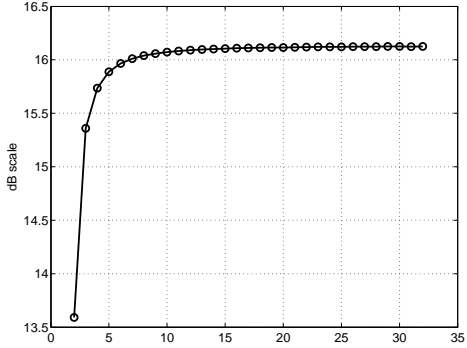}
\caption{ZZB threshold SNR $\texttt{SNR}_{\text{th}}$ for estimating the frequency of a $N_{t,1D} \times N_{t,1D}$ sinusoid as a function of $N_{t,1D}$}
\label{fig:threshold_SNR}
\end{figure}

The preceding criteria delineate the regime in which the \emph{geometry} of the estimation problem is preserved approximately. We now turn to another factor which affects estimation performance, namely the SNR. Consider measurements of the form
\begin{equation}\label{eq:measurements_ZZB}
y_{m,n} = e^{j\left(\omega_1 m + \omega_2 n + \phi\right)} + z_{m,n},~0\leq m,n\leq N_{t,1D}-1,
\end{equation}
where $z_{m,n}$ are i.i.d. $\mathcal{CN}(0,\sigma^2)$ and spatial frequencies $\omega_1, \omega_2$ and phase $\phi$ are parameters to be estimated. The Cram\'er Rao Bound\cite{VanTrees:Bell:Book:Bounds} (CRB) for estimating $\omega_1$ from measurements \eqref{eq:measurements_ZZB} is given by $C(\sigma^2) = \fracc{6}{\left(\texttt{SNR}\left(N_{t,1D}^2-1\right)\right)}$, where $\texttt{SNR} = \fracc{\norm{\bx_t(\bomega)}^2}{\sigma^2} = \fracc{N_{t,1D}^2}{\sigma^2}$ (same expression holds for $\omega_2$). Assuming an uniform prior over $[0,2\pi)^3$ for the parameters $(\omega_1, \omega_2, \phi)$, the Ziv Zakai Bound (ZZB) with periodic distortion\cite{Basu::ZZB::Periodic::2000} for estimating $\omega_1$ evaluates to:
$$
Z(\texttt{SNR}) = \int_0^\pi Q\left( \sqrt{ \texttt{SNR} \left(1 - \abs{\frac{\sin\left(N_{t,1D}h/2\right)}{N_{t,1D}\sin\left(h/2\right)}}\right)} \right) h\text{d}h.
$$
An indicator of the $\texttt{SNR}$ needed for successful estimation is the convergence of the ZZB to the CRB\cite{Ramasamy_TSP_2014}. We use the $\texttt{SNR}$ beyond which the ZZB is within $0.1$dB of the CRB as a measure of this convergence. We plot this ZZB \emph{threshold} SNR for different values of $N_{t,1D}$ in Figure~\ref{fig:threshold_SNR}. e.g, $\texttt{SNR}_\text{th} = 16.04$dB for an $8 \times 8$ array and $\texttt{SNR}_\text{th} = 16.13$dB for a $32 \times 32$ array. 

The total energy $\texttt{E}_\text{tot}$ corresponding to the $i$-th path collected across the $ML$ measurements $\{\by_j,1\leq j\leq L\}$ is given by:
\begin{align*}
\texttt{E}_\text{tot} & = \norm{\bA\bx_t\left(\bomega_i^t\right)}^2 \times \sum_{j = 1}^{j = L} \abs{h_{i,j}}^2.
\end{align*}
Using \eqref{eq:relate_h_B} in the above, we have that
\begin{align*}
\texttt{E}_\text{tot} & = \norm{\bA\bx_t\left(\bomega_i^t\right)}^2 \norm{\bB\bx_r\left(\bomega_i^r\right)}^2 {P_e}\abs{g_i}^2\\
&\approx M L N_{t,1D}^2 N_{r,1D}^2 {P_e} \abs{g_i}^2\\
& = M L P N_{r,1D}^2 \abs{g_i}^2,
\end{align*}
where $P = N_{t,1D}^2 P_e$ is the total transmit power supplied to the $N_{t,1D} \times N_{t,1D}$ antenna array. The above approximation holds when $M$ and $L$ satisfy the preceding geometry preservation criteria in Sections~\ref{sub-sec:geometry_M}~and~\ref{sub-sec:geometry_L} respectively. The effective SNR of the $i$-th path is given by $ \texttt{SNR}_\text{eff} = \fracc{\texttt{E}_\text{tot}}{\sigma^2}$. It is important to note that the per-measurement noise variance $\sigma^2$ is given by $\sigma^2 = N_{r,1D}^2 \sigma^2_{e}$, where $\sigma_e^2$ is the noise variance per antenna element. Assuming no interference (which we account for in Section~\ref{sec:reuse}), $\sigma_e^2 = \text{N}_0 W_s$ with $W_s$ denoting the sounding bandwidth and $\text{N}_0$ the thermal noise floor. The reason for the scale factor $N_{r,1D}^2$ in the expression for $\sigma^2$ is the following: Our measurement process consists of multiplying the received signal at each antenna (of which there are $N_{r,1D}^2$) by phasors $b(m,n,k)\in\{\pm 1,\pm j\}$ and \emph{adding} the resultant signal (as per \eqref{eq:receive_antenna-construction}). Since thermal noise seen by the $N_{r,1D}^2$ isolated receive antennas are independent random variables, we have that $\sigma^2 = N_{r,1D}^2 \times \left(\text{N}_0 W_s\right)$. Therefore, the effective SNR of the $i$-th sinusoid is given by:
$$
\texttt{SNR}_\text{eff} = ML P \fracc{\abs{g_i}^2}{\text{N}_0 W_s}.
$$
This must exceed the ZZB threshold $\texttt{SNR}_{\text{th}}$ for successful estimation. Noting that $\fracc{ML}{W_s}$ is the time taken for channel sounding, the ZZB threshold $\texttt{SNR}_\text{th}$ gives us the means to evaluate the minimum overhead in time to estimate the channel for a given path gain $\abs{g_i}^2$:
\begin{equation}\label{eq:sounding_time}
\text{Time taken} = \fracc{ML}{W_s}   \geq \fracc{ \texttt{SNR}_\text{th} ~ \text{N}_0}{P \abs{g_i}^2}
\end{equation}
The size of the picocell gives us a lower bound on $\abs{g_i}^2$ and we later use this to guide us in choosing the sounding bandwidth $W_s$ using \eqref{eq:sounding_time}.

\subsection{Sounding rate}

\begin{figure}
\centering
\includegraphics[width=0.7\columnwidth]{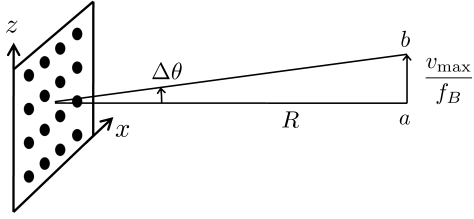}
\caption{Geometry corresponding to maximum change in $\omega_z$: The user moves along the $z$ axis at a speed of $v_{\max}$ in the time interval $\fracc{1}{f_B}$ between two consecutive channel sounding rounds}
\label{fig:how_often_sound}
\end{figure}

We round off the discussion on choice of protocol parameters by giving a rule of thumb for the rate/frequency $f_B$ at which the spatial channel $\{\bomega_i\}$ needs to be \emph{reestimated}. We use the estimated spatial frequency $\hat{\bomega}$ for beamforming purposes in the time period between two channel sounding rounds (communication phase sandwiched between consecutive sounding phases; see Figure~\ref{fig:sound_and_comm}). Following the discussion in Section~\ref{sub-sec:algo-tracking}, we have that if $\norm{\bomega(t) - \hat{\bomega}}_\infty < \fracc{\pi}{N_{t,1D}}$ throughout the communication phase, where $\bomega(t)$ denotes the true spatial frequency and $\hat{\bomega}$ the estimate from the prior sounding round, then the \emph{loss} in SNR, given by $\fracc{\norm{\bx_t(\bomega(t))}^2}{\innerProd{\bx_t\left(\hat{\bomega}\right)}{\bx_t\left({\bomega(t)}\right)}}$, is smaller than $3$dB. If we assume that the closest user to the basestation array is at a distance $R$ meters and that the maximum speed of a user in the picocell is given by $v_{\max}$ meters per second, then the maximum change (in terms of the $\ell_\infty$-norm) in spatial frequency $\Delta\omega$ between consecutive sounding phases, spaced $\fracc{1}{f_B}$ apart, is given by $\fracc{2\pi d v_{\max}}{f_B R \lambda}$. The worst-case geometry which achieves this bound is when the user is at a distance $R$ along the bore-sight of the array and heading in a direction aligned with the one of the array axes. For this worst-case geometry (plotted in Figure~\ref{fig:how_often_sound}), we have that:
\begin{align*}
\Delta\omega & \leq  (\fracc{2\pi d}{\lambda})\sin\Delta\theta  \approx \frac{2\pi d v_{\max}}{f_B \lambda  R}.
\end{align*}
Assuming that the estimate $\hat{\bomega}$ from the previous sounding phase is accurate, if we ensure that $\fracc{2\pi d v_{\max}}{f_B R \lambda} \leq \fracc{\pi}{N_{t,1D}}$, we have that the beamforming losses in the intervening period are smaller than $3$dB. This tells us that channel needs to be sounded often enough so that
\begin{equation}\label{eq:sounding_rate}
f_B \geq \fracc{ 2  d  v_{\max} N_{t,1D}}{  R \lambda}.
\end{equation}
In the following discussions, we use the preceding in conjunction with \eqref{eq:sounding_time} to determine the overhead incurred in estimating the channel using the compressive architecture proposed herein.

\section{System design}\label{sec:link_budget}

We now discuss some key aspects of downlink system design related to our compressive architecture. We start by choosing basestation transmit power based on rules set by regulatory authorities, and then filling in the other details of the protocol according to the prescriptions laid out in Section~\ref{sec:protocol_choice}. Fixing the mobile array to be $4 \times 4$ ($N_{r,1D} = 4$), we consider two different choices for the base station array size: $8 \times 8$ ($N_{t,1D}  = 8$) and $32 \times 32$ ($N_{t,1D} = 32$). All arrays are $d = \lambda/2$-spaced. The total available bandwidth for communication and sounding is $2$GHz. 

\subsection{Transmit power}
We fix the effective isotropically radiated power (EIRP) to $40$dBm, consistent with Federal Communications Commission (FCC) regulations for 60 GHz unlicensed transmission. Accounting for transmit beamforming gain using an $N_{t,1D} \times N_{t,1D}$ array, the total transmit power
\begin{equation} \label{eq:transmit_power}
P = 40 - 20 \log N_{t,1D}~\text{dBm}
\end{equation}
which evaluates to $22$dBm and $10$dBm for $N_{t,1D} = 8$ and $N_{t,1D} = 32$ respectively. Assuming that this power is evenly split among the $N_{t,1D}^2$ transmit elements, the power per transmit element is given by $P_e = P - 20 \log N_{t,1D}~\text{dBm} = 40 -20 \log N_{t,1D}~\text{dBm}$, which evaluates to $4$dBm and $-20$dBm, respectively. Assuming that we design each element to operate at a fixed power, this is also the power per element used in the beaconing phase, even though the latter does not get the benefit of transmit beamforming.

\subsection{Communication range}

In order to ensure that the SNR for compressive estimation is adequate over a picocell, we first determine the picocell size using a nominal communication link budget, and then calculate the overhead required for successful estimation at that range. Standard link budget calculations, assuming oxygen absorption of 16 dB/km, an EIRP of $40$dBm and a $4 \times 4$ receive array providing directivity gains of 12 dBi, can be used to show that we can attain a per-symbol SNR of $6$dB at a link margin of $10$dB for a symbol rate of $2$GHz at a range of $200$m. For omnidirectional free space propagation, the power gain in dB as a function of range $r$ is given by
$$
G_{dB} (r) =- \mu r +  20 \log_{10} \frac{\lambda}{4 \pi r} ~~{\rm (dB)}
$$
where $\mu = 0.016$dB/m to account for oxygen absorption. Note that $\mu$ can be increased in order to account for rain. However, since our purpose is to ensure that channel estimation is successful whenever communication is successful, the contribution due to $G_{dB} (r)$ cancels out, as we show shortly. Thus, while the particular value of $\mu$ determines picocell size, we shall see that it does not affect the overhead for channel estimation.

The SNR per symbol is given by
\begin{align}  \label{eq:snrc}
\texttt{SNR}_{\text{c}} {\rm (dB)} &= {\rm EIRP ~(dBm)} + G_{dB} (r) + 20 \log_{10} N_{r,1D} \\ \nonumber
& - 10\log_{10} \left(\text{N}_0 W_c\right) - L_{margin} (comm)
\end{align}
where $L_{margin} (comm)$ is the link margin (dB) for communication. Note that $10 \log_{10} N_0 = -174 + NF$ dBm over a bandwidth of 1 Hz, where $NF$ denotes the receiver noise figure in dB. Plugging in $W_c = 2$GHz, $N_{r, 1D} = 4$, and $NF=6$dB, we obtain a per symbol SNR of 7 dB at a range of $r=100$ meters.  

\subsection{Channel sounding protocol}\label{sub-sec:sounding-protocol}
\begin{figure}
\centering
\includegraphics[width=0.95\columnwidth]{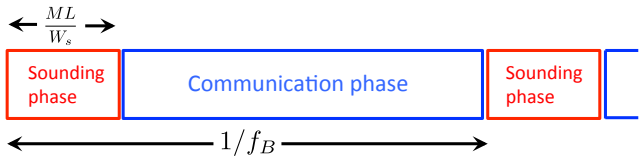}
\caption{Channel sounding and communication phases of the proposed system} 
\label{fig:sound_and_comm}
\end{figure}

Our channel sounding protocol is specified by four parameters: (i) bandwidth used by each basestation when sounding the channel, which we denote by $W_s$ (ii) number of transmit beacons (or virtual transmit antennas) $M$ (iii) number of receive measurements per transmit beacon (or virtual receive antennas) $L$ and (iv) sounding rate $f_B$ which determines how often the channel is sounded. The parameters $M,L$ and $W_s$ together determine the effective sounding SNR. This must exceed the ZZB threshold SNR for successful channel estimation. This gives rise to the condition in \ref{eq:sounding_time}. Imposing an estimation link margin $L_{margin} (est)$ (dB) and going to the dB domain, we have
\begin{align} \label{eq:sounding_time_with_distance}
10\log_{10} \left(\fracc{ML}{W_s}\right)  &  \geq  \texttt{SNR}_{\text{th}} + L_{margin} (est) \\
& +10 \log_{10} N_0 - P - G_{dB} (r) \nonumber
\end{align} 
Adding (\ref{eq:sounding_time_with_distance}) and (\ref{eq:snrc}) and simplifying, we obtain
\begin{align*}
10\log_{10} \left(\fracc{ML}{W_s}\right)  &\geq \texttt{SNR}_{\text{th}} - \texttt{SNR}_{\text{c}} + L_{margin} (est) \\
&- L_{margin} (comm) + 20 \log_{10} N_{t,1D} \\
& + 20 \log_{10} N_{r,1D} - 10 \log_{10} W_c
\end{align*}
The key take-away is that $ML/W_s$ must be large enough to compensate for the fact that we do not have the benefit of beamforming during the sounding phase. Notice that the range $r$ (i.e., the dependence on picocell size) has cancelled out. Setting  $L_{margin} (est)= 16$dB (we use a higher link margin for channel sounding to account for power losses due to randomness of $\bA$ and $\bB$),
we obtain
\begin{equation}\label{eq:sounding_time_with_power}
\text{Time taken} = \frac{ML}{W_s} \geq  \begin{cases} 16.34~\mu \text{s} & N_{t,1D}=8\\ 0.2669 ~\text{ms}& N_{t,1D}=32. \end{cases}
\end{equation}
We choose the number of transmitter beacons for the $8 \times 8$ and $32 \times 32$ transmitter arrays based on the geometry preservation criterion for the transmitter's spatial channel estimation problem discussed in Section~\ref{sub-sec:geometry_M}. We use $M=24$ for $N_{t,1D} = 8$ and $M = 30$ for $N_{t,1D} = 32$ by numerically evaluating the worst-case distortion of pairwise distances relevant for the channel estimation problem (in Figure~\ref{fig:choice_of_M} we plot the worst-case distortion as a function of $L$ for a random instance of $\bA$ and $N_{t,1D} = 32$). Using the receive energy preservation criterion given in Section~\ref{sub-sec:geometry_L}, we choose the number of receive weights for the $4\times 4$ receive array as $L = 6$. Using these values for $M$ and $L$ in \eqref{eq:sounding_time_with_power}, we obtain that the channel sounding bandwidth must satisfy
$$
W_s \leq \begin{cases} 8.8124~\text{MHz} & N_{t,1D}=8\\  674.34~\text{KHz} & N_{t,1D}=32. \end{cases}
$$
We choose $W_s =  8.8124$ MHz for $N_{t,1D}=8$ and $W_s = 674.34$ KHz for $N_{t,1D}=32$ so as to minimize the overhead in time, given by $\fracc{ML}{W_s}$. Our specification of the channel sounding protocol will be complete when we give $f_B$, the rate at which we sound the channel (see Figure~\ref{fig:sound_and_comm}) which must satisfy \eqref{eq:sounding_rate}. Assuming that the closest user is at  a distance of $R = 20$m and that the maximum speed of a user in the picocell $v_{\max}$ is $45$ miles per hour ($20$ m/s), we have that: $f_B \geq 8$ Hz for $N_{t,1D}=8$ and $f_B \geq 32$ Hz for $N_{t,1D}=32$. Choosing the minimum value for $f_B$, we have that the overhead for our channel sounding protocol is $\fracc{ML f_B}{W_s} = 0.0131\%$ for $N_{t,1D}=8$ and $0.8542 \%$ for $N_{t,1D}=32$.

\subsection{Reuse analysis for channel sounding}\label{sec:reuse}

\begin{figure}
\centering
    \includegraphics[width=3.2in]{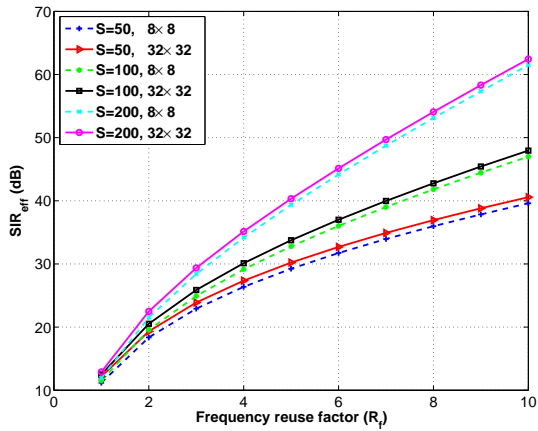}
\caption{Effective Signal to Interference Ratio ${\texttt{SIR}_{\text{eff}}}$ for $ML = 24 \times 6$ ($8 \times 8$ scenario) and $ML = 30 \times 6$ ($32 \times 32$ scenario; dashed) as a function of reuse factor $R_f$}
\label{fig:FR}
\end{figure}

We investigate how a sequence of basestations employed in an urban canyon environment can share resources when estimating the spatial channel to users in their respective cells. The envisioned mm-wave system involves alternating between channel estimation and communication phases as shown in Figure~\ref{fig:sound_and_comm}. We assume that channel sounding rounds across basestations are aligned in time. We now characterize how the $2$GHz spectrum is to be shared in space so as to limit the effect of interference from neighboring picocells on channel estimation performance. Such interference management is essential in the sounding phase; unlike the highly directive beams used in the communication phase, compressive sounding beacons are essentially omnidirectional. To see this, consider the average transmit power along any direction $\bomega$. This is given by $P_e\fracc{\norm{\bA\bx_t(\bomega)}^2}{L}\approx P_e \norm{\bx_t(\bomega)}^2 = P_e N_{t,1D}^2 = P$, the total transmit power. The approximation $\fracc{\norm{\bA\bx_t(\bomega)}^2}{L}\approx {\norm{\bx_t(\bomega)}^2}$ holds when the number of beacons $L$ is large enough. Therefore, the average energy \emph{per-measurement} received by an antenna at a distance $r$ from a transmitter sending compressive beacons is given by $P G(r)$, where $G(r) = 10^{G_{dB}(r)/10} = \fracc{\lambda^2}{\left(16 \pi^2 r^2\right)} e^{- \nu r}$ ($\nu = \left(\fracc{\mu}{10}\right) \ln 10$) is the omnidirectional power gain at range $r$. We assume that basestations are deployed regularly as shown in Figure~\ref{fig:FR} and that the inter-basestation separation (along the street) is given by $S$. Suppose that the reuse factor is $R$ (i.e, every $R$th basestation uses the same slice of the frequency spectrum to estimate downlink spatial channels). We assume that for narrow urban canyons, the distance between a user and all interfering basestations (those that are allocated the same sounding BW) are well approximated by $\{kR_f S,~ k\in \mathbb{Z}\setminus\{0\}\}$. Thus, the interference power seen by a single antenna is given by
\begin{equation*}
I = 2 \times \sum_{k=1}^{k=\infty} \sum_{\text{paths}} P G(k R_f S) = 8 P \sum_{k=1}^{k=\infty} G(k R_f S),
\end{equation*}
where we have assumed that there are 4 viable paths between the interfering basestation and user, each introducing the same amount of interference as the LoS path. This is a pessimistic assumption, since NLOS paths are attenuated by larger path lengths and reflection losses. Plugging in the expression for $G(r)$, we have that
\begin{align*}
I &= \left(\fracc{P \lambda^2 }{2 \pi^2 R_f^2S^2}\right)  \sum_{k=1}^{k=\infty} \fracc{e^{-\nu R_f S k}}{k^2}\\
& = \left(\fracc{P  \lambda^2 }{2 \pi^2 R_f^2S^2}\right) \text{Li}_2\left(e^{-\nu R_f S}\right),
\end{align*}
where $\text{Li}_2(z) = \sum_{k=1}^{k=\infty} \fracc{z^k}{k^2}$ is the dilogarithm function. The interference seen per antenna adds to thermal noise to give an effective per-element noise level of $\sigma_e^2 = \text{N}_0 W_s + I$. Assuming a worst-case geometry for the user of interest (distance of $S$ from the basestation) and proceeding as in Section~\ref{sub-sec:SNR_needed}, we see that effective Signal to Interference and Noise Ratio $\texttt{SINR}_{\text{eff}}$ is given by
$$
\texttt{SINR}_{\text{eff}} = \fracc{MLP G(S)}{\sigma_e^2}.
$$
This can be rewritten as
$$
\fracc{1}{\texttt{SINR}_{\text{eff}}} = \fracc{1}{\texttt{SNR}_{\text{eff}}} + \fracc{1}{\texttt{SIR}_{\text{eff}}},
$$
where $\texttt{SNR}_{\text{eff}} =\fracc{MLP G(S)}{\text{N}_0 W_s}$ and the Signal to Interference Ratio $\texttt{SIR}_{\text{eff}} =\fracc{MLP G(S)}{I} = \fracc{ML R_f^2 e^{-\mu S}}{8 \text{Li}_2\left(e^{-\mu R_f S}\right)}$. We need to ensure that $\texttt{SINR}_{\text{eff}}$ exceeds the ZZB SNR threshold for successful estimation. We choose the reuse factor $R_f$ so that we are in the noise-limited regime by setting
$$
\fracc{1}{\texttt{SIR}_{\text{eff}}} < 0.1 \times \left(\fracc{1}{\texttt{SNR}_{\text{th}}}\right) \approx -10 - 16~\text{dB}.
$$
Assuming that protocol parameters are chosen so that $\texttt{SNR}_{\text{eff}}$ exceeds $\texttt{SNR}_{\text{th}}$, we can ignore interference in $\texttt{SINR}_{\text{eff}}$ calculations when 
$$
\texttt{SIR}_{\text{eff}} > \texttt{SNR}_{\text{th}} + 10 \approx 26\text{dB for }N_{t,1D}=8,32.
$$
In Figure~\ref{fig:FR}, we plot achievable effective SIRs as a function of frequency reuse factor $R_f$ for two example systems in Section~\ref{sub-sec:sounding-protocol}: i.e, $8 \times 8$ and $32 \times 32 $ arrays with total number of measurements given by $ML = 24\times 6$ and $ML = 30 \times 6$ respectively. As the picocell size $S$ grows, exponential attenuation due to oxygen absorption (the $e^{-\nu S}$ term in the expression for $\texttt{SIR}_{\text{eff}}$) helps in attenuating interference and improving SIR for same reuse factor $R_f$. To illustrate this we plot SIR as a function of $R_f$ for three cell sizes $S = 50,100,200$m in Figure~\ref{fig:FR}. We observe that, in order to ensure $\texttt{SIR}_{\text{eff}} > 26$dB, a reuse factor of $R_f = 4$ is needed for $S=50$m, while $R_f=3$ suffices for $S=200$m. Plugging in the per-basestation sounding bandwidth $W_s$ calculations in Section~\ref{sub-sec:sounding-protocol}, we see that the overall \emph{system-level} channel sounding bandwidth $W_s \times R_f$ for a $S=50$m picocell is as small as $8.8124\text{MHz} \times 4 = 35.2$MHz and $674.34\text{KHz} \times 4 = 2.7$MHz for $N_{t,1D} = 8, 32$, respectively, which is dwarfed by the total available bandwidth ($2$GHz).  

\begin{figure}
\centering
    \includegraphics[width=0.9\columnwidth]{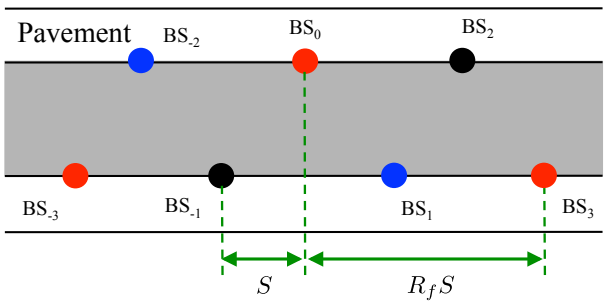}
\caption{Reuse of frequency resources for reuse factor $R_f = 3$}
\label{fig:cells}
\end{figure}

\section{Simulation results}\label{sec:sims}

We perform simulations for the two example systems considered in Section~\ref{sec:link_budget} ($8 \times 8$ and $32 \times 32$ transmit arrays). We report results for the algorithm proposed in Section~\ref{sec:algo} and two feedback strategies: (i)  `full': users feedback the measured virtual channel matrix $\bY$ ($M \times L$ matrix; $L=6$ for both systems) and (ii) `svd': users feedback the $2$ dominant left singular vectors of $\bY$, scaled by their corresponding singular values ($M \times 2 $ matrix; one-third feedback overhead).

We consider $6$ mobile users moving in the urban canyon at speeds of $20$, $3$, $15$, $1.5$, $2.1$ and $10$ meters per second (covering both vehicular and pedestrian settings). The height at which each mobile device is held is in the $1.3 - 1.4$m range. The basestation is mounted on a lamppost on the pavement ($7$ meters from a canyon wall), at a height of $6$ meters. The basestation antenna array is tilted by about $7.5^{\circ}$ in both the azimuth and elevation directions so that the boresight of the array points towards middle of the corresponding cell. This helps in more accurate spatial frequency estimation: since a change in direction near the boresight of the array results in larger changes in spatial frequencies than far away from the boresight, resolving paths is easier when the array points towards a direction in which we are likely to see more paths.  We do not model blockage in these simulations, assuming that the LoS path and the three first order reflections are all available. Our goal is to estimate and track the $K=4$ paths to all $6$ users.

\begin{figure}
\centering
\includegraphics[width=0.7\columnwidth]{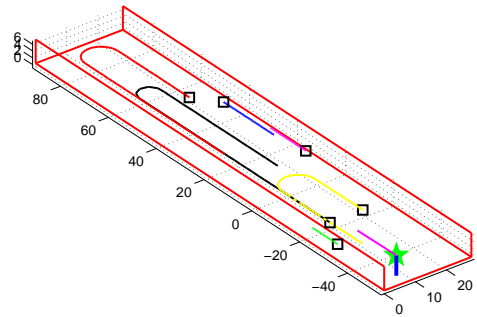}
\caption{Six users in the urban canyon moving over the duration of the $7$ second simulation interval. Their positions at time $t = 0$ is marked using a $\bm{\square}$-symbol} 
\label{fig:sim_scenario}
\end{figure}

\noindent\textbf{Estimation error:}~Let $T = \{\bomega_m~:~ m =1,\dots,K\}$ denote the true spatial frequencies and $P = \{\hat{\bomega}_n~:~n=1,\dots,\hat{K}\}$ denote the set of estimated spatial frequencies. When the base station uses one of the estimates in $P$, say $\hat{\bomega}$, to form a beam, we do not realize the full $20\log N_{t,1D}$ dB beamforming gain. A measure of the sub-optimality is the estimation error $\norm{\bomega - \hat{\bomega}}_2$, which we normalize by the DFT spacing of $\fracc{2 \pi}{N_{t,1D}}$ to define the following error metric:
\begin{equation}\label{eq:est_errors}
\Delta\omega(m) = \fracc{\min_n \norm{\bomega_m - \hat{\bomega}_n}_2}{\left(\fracc{2 \pi}{N_{t,1D}}\right)}.
\end{equation}
When no true spatial frequency exists near an estimate $\hat{\bomega}$, i.e, when $\hat{\bomega}$ is a ``phantom estimate'', we will quickly be able to discard it when we beamform in the direction of $\hat{\bomega}$ and find that the mobile does not receive power commensurate to what it expects with the $20\log N_{t,1D}$ beamforming gain. 

\begin{figure*}[t!]
\centering
\begin{subfigure}[t]{0.48\textwidth}
        \centering
	\includegraphics[width=0.49\columnwidth]{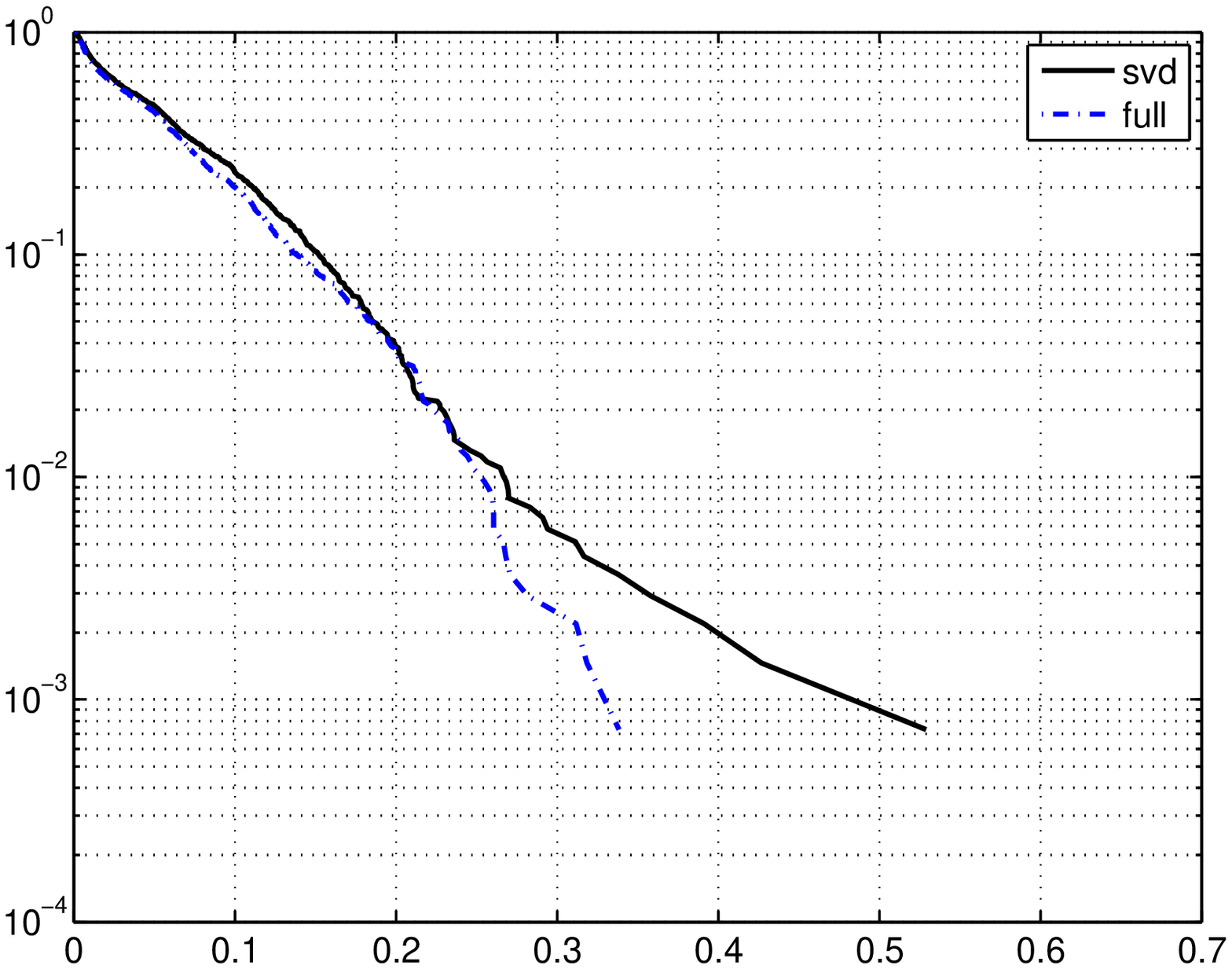}~					\includegraphics[width=0.49\columnwidth]{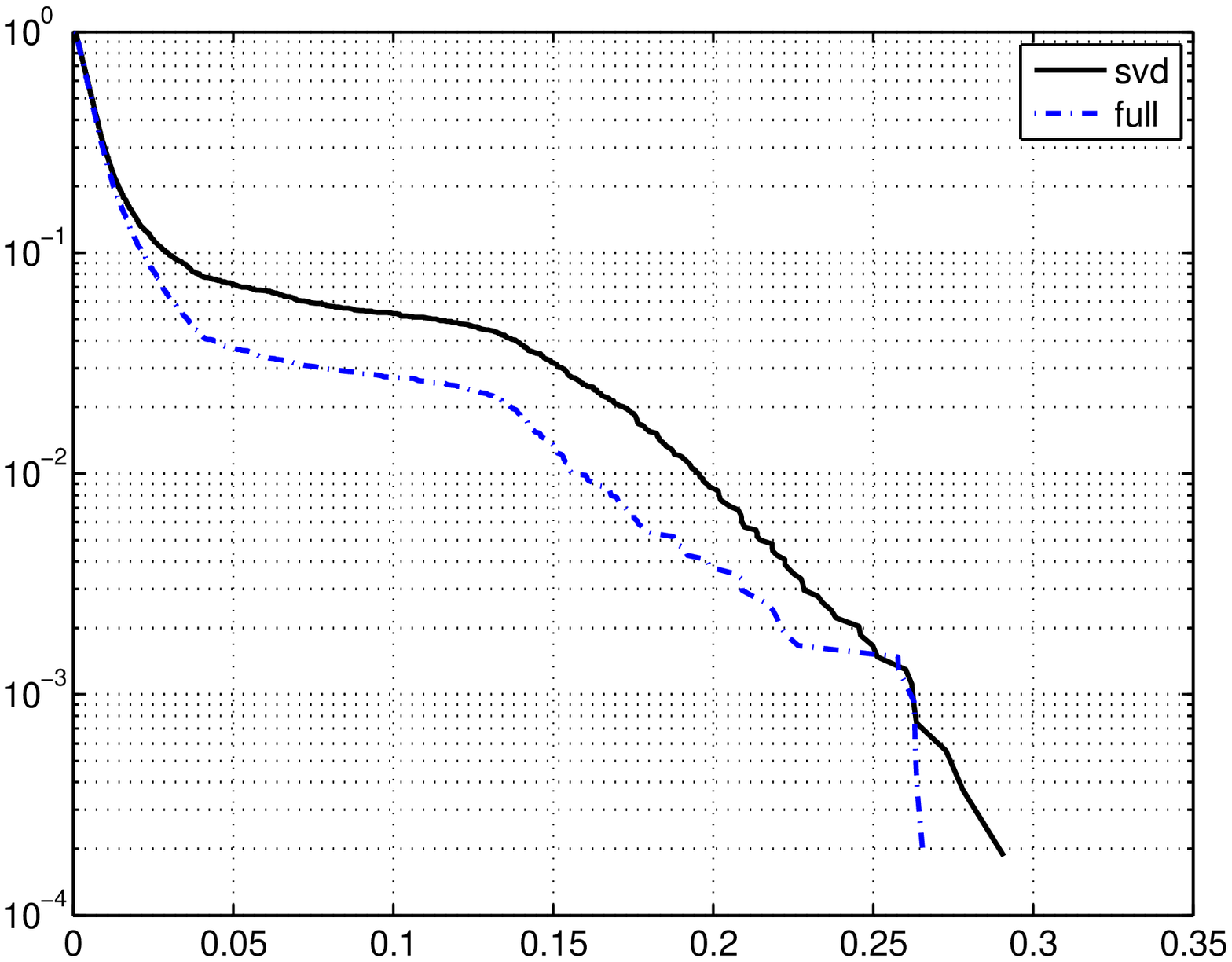}
	\caption{CCDF of frequency estimation errors \eqref{eq:est_errors} for $N_{t,1D} = 8$ (left) and $N_{t,1D} = 32$ (right)}
	\label{fig:sim_errors}

\end{subfigure}
~~
\begin{subfigure}[t]{0.48\textwidth}
	\centering
	\includegraphics[width=0.485\columnwidth]{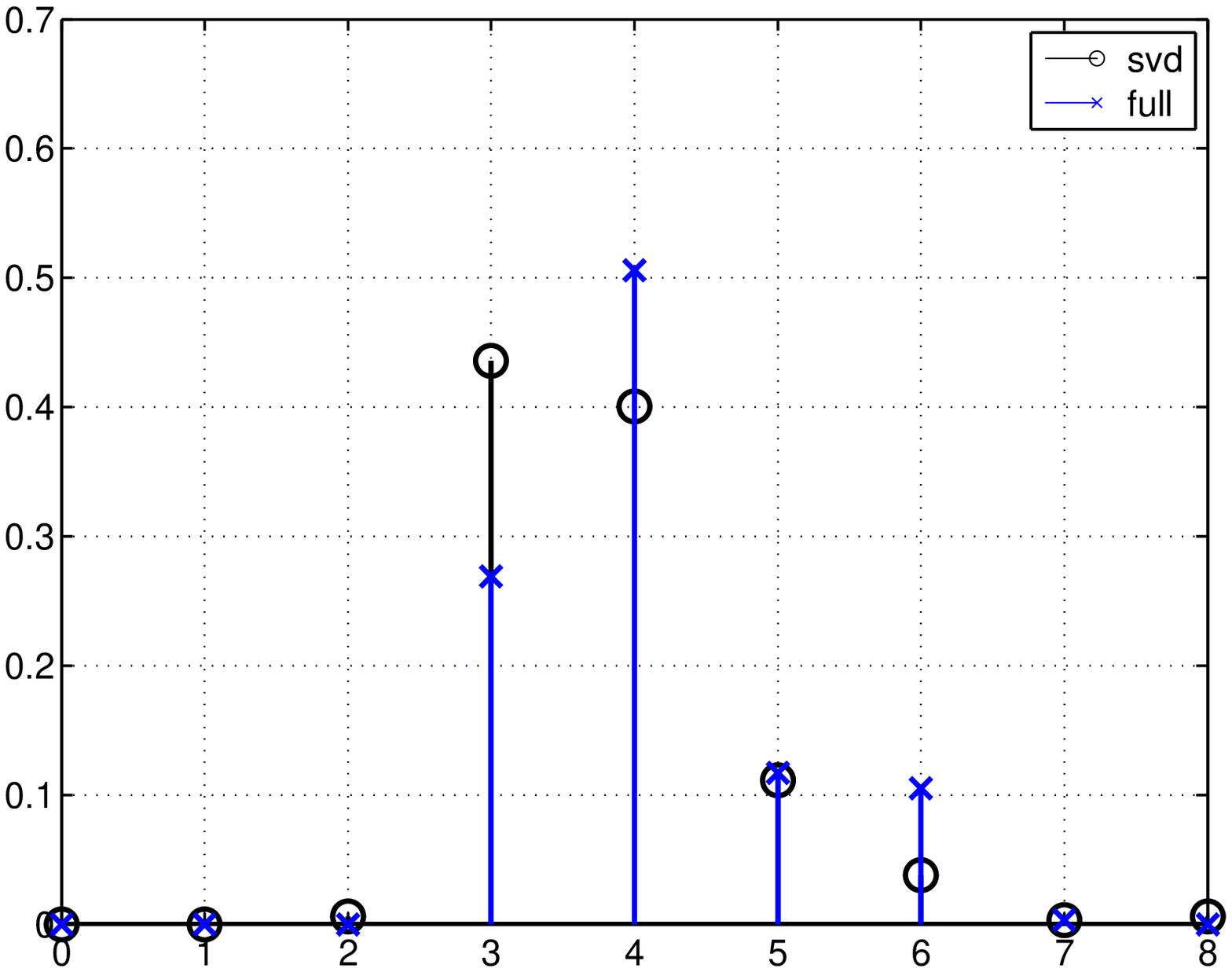}~			\includegraphics[width=0.485\columnwidth]{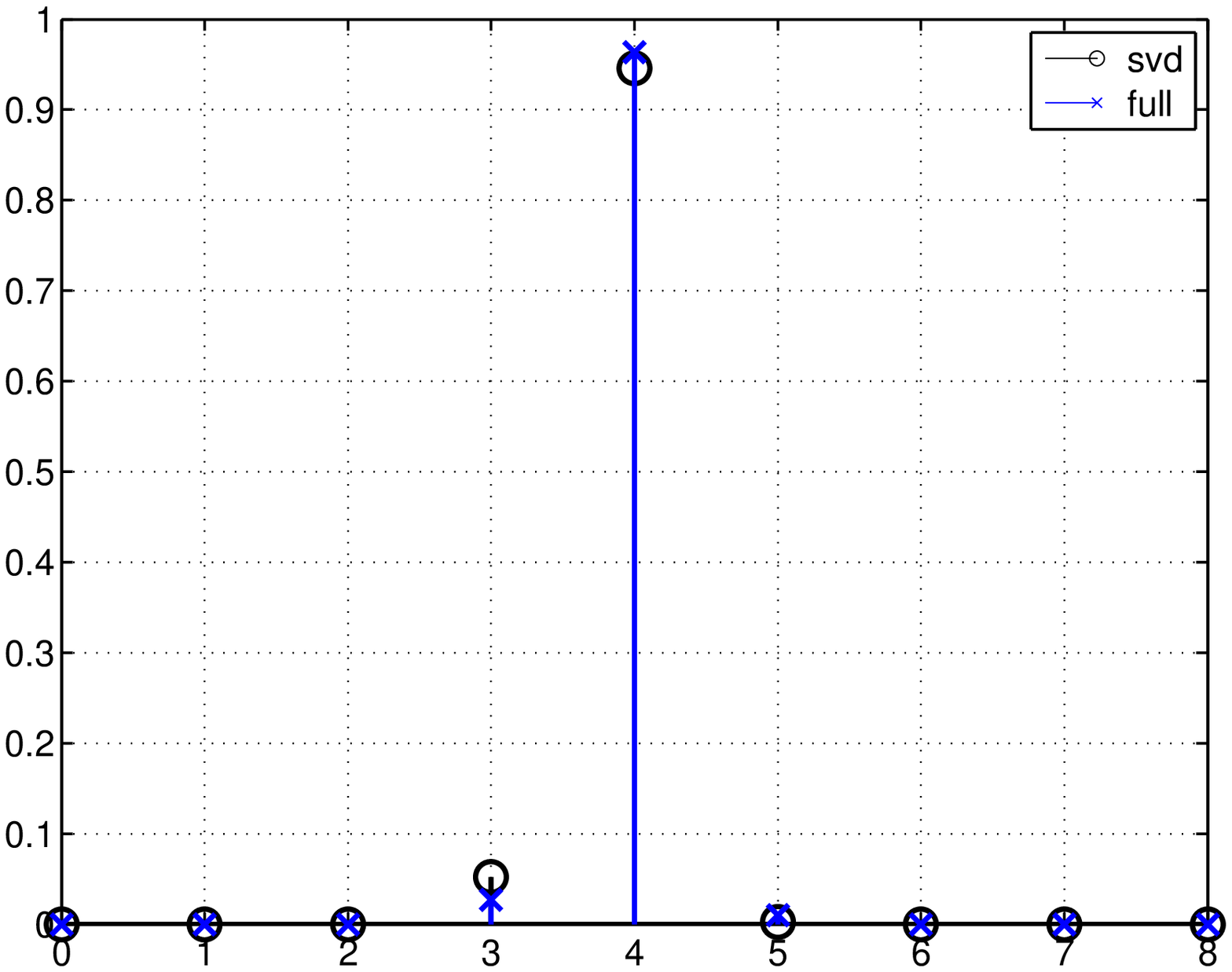}
	\caption{PDF of \# of estimated paths $\hat{K}$ ($K = 4$) for $N_{t,1D} = 8$ (left) and $N_{t,1D} = 32$ (right)} .
	\label{fig:sim_errors_support_size}
\end{subfigure}
\caption{Two feedback strategies considered: (i) $M \times 6$ matrix $\bY$ (`full') and (ii) top two dominant singular vectors (one-third overhead)}
\label{fig:sims}
\end{figure*}

We plot the Complementary Cumulative Distribution Function (CCDF) of estimation errors \eqref{eq:est_errors} for the two systems ($N_{t,1D} = 8, 32$) in Figure~\ref{fig:sim_errors} and the Probability Distribution Function (PDF) of the number of paths estimated $\hat{K}$ (correct value is $K = 4$) in Figure~\ref{fig:sim_errors_support_size}. From Figure~\ref{fig:sims}, we see that feedback of dominant singular vectors is an efficient feedback strategy which performs just as well as feeding back the entire matrix $\bY$, while using only a third of uplink resources.

\begin{figure}
 \begin{center}
    \includegraphics[width=3.2in]{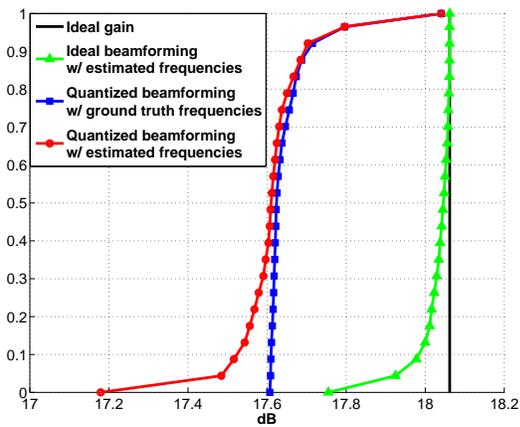}
\caption{Beamforming gain achieved by an $8\times8$ antenna array for ideal and quantized beamforming techniques}
\label{fig:BeamformingGain8x8}
\end{center}
\end{figure}

Next, in order to evaluate the effect of errors in spatial frequency estimation on beamforming performance, we simulate a simple scenario in which the transmitter beamforms toward the strongest estimated path. Figure \ref{fig:BeamformingGain8x8} shows the CDF of the achievable beamforming gain for an $8 \times 8$ array. While ideal beamforming requires adjustment of both gains and phases, suboptimal approaches for RF beamsteering with severely quantized phase-only control (four phases) have been studied in our earlier conference paper \cite{ramasamy_ita12}. We see from Figure \ref{fig:BeamformingGain8x8} that if ideal beamforming were performed with our estimates, then the SNR loss is less than 0.3 dB.  If four-phase control is used based on our estimates, then the SNR loss is less than 1 dB. The results for $32 \times 32$ arrays are entirely similar, and are therefore not plotted here..

\begin{figure}
\centering
    \includegraphics[width=1.7in]{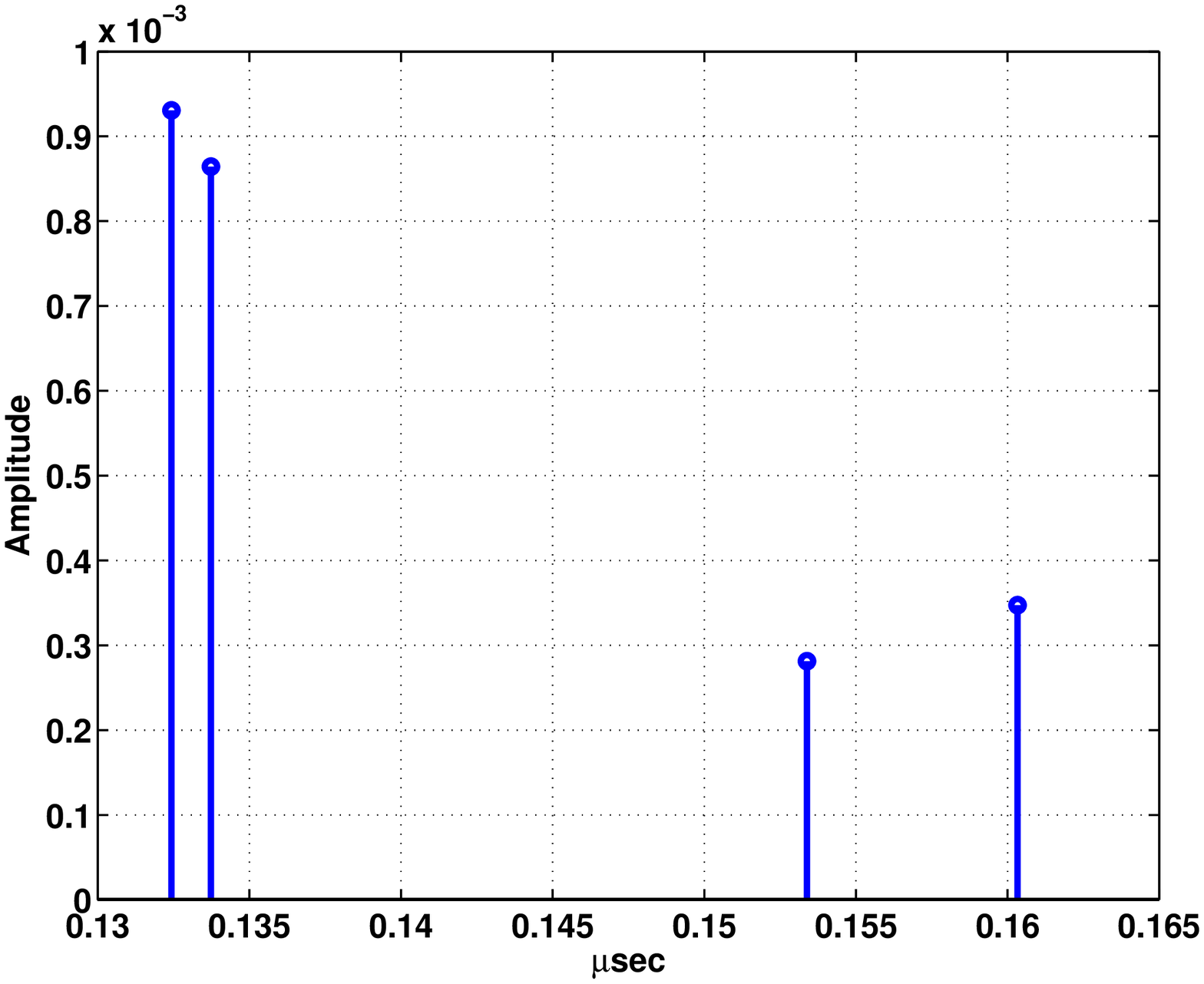}~\includegraphics[width=1.7in]{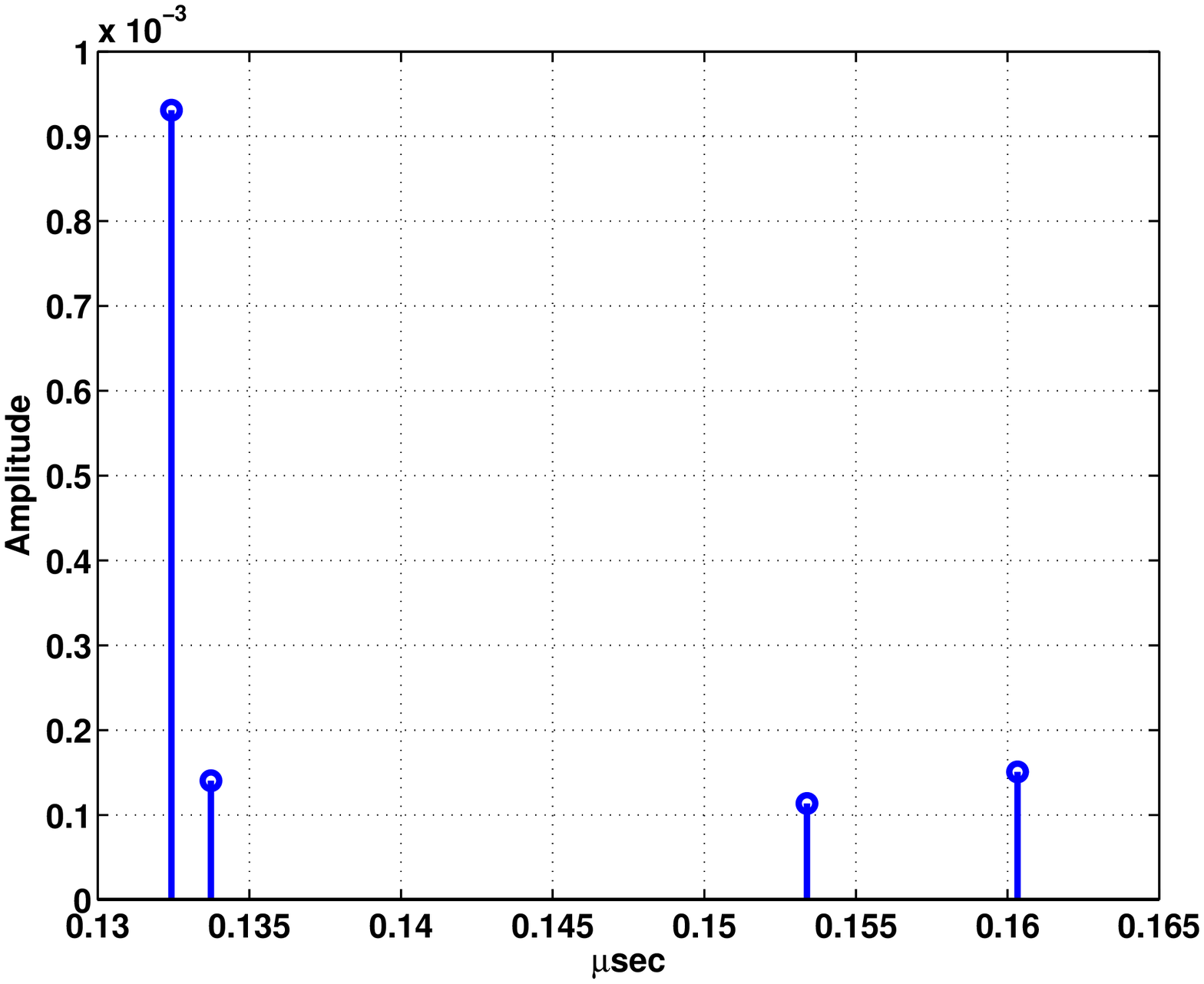}
    
\caption{Channel impulse response with quantized beamforming towards estimated strongest path for the $8 \times 8$ (left) and $32 \times 32$ (right) scenarios}
\label{fig:Impulseresponse}
\end{figure}

Thus far, we have not said anything about channel frequency selectivity.  Our proposed algorithm uses a small segment of the band to estimate the spatial channel, and the problem of channel dispersion is not addressed.  However, we note that beamforming using a large array should reduce the effect of undesired paths, which simplifies the task of equalization. Figure \ref{fig:Impulseresponse} shows the channel impulse responses for the $32\times 32$ and $8 \times 8$ antenna arrays for a typical snapshot, when the transmitter beamforms towards the strongest estimated path. In our simulated setting, the LoS and ground reflection are close to each other in terms of both delays and angles of departure. We see that $8\times 8$ array fails to resolve them, with both paths falling into the antenna's main lobe, while the $32 \times 32$ antenna array, which has smaller beamwidth ($4^{\circ}$ half power beamwidth), attenuates the undesired tap down to one-ninth of the desired path. Of course, it is possible to utilize the channel estimates far more intelligently, potentially with nulls directed both at strong undesired paths for the mobile of interest, and at the dominant paths for other nearby mobiles. The latter can be particularly useful for combating intra-cell interference when a base station face has multiple antenna arrays, each communicating with a different mobile.

\section{Conclusions} \label{sec:conclusions}

We have shown that it is possible to super-resolve mm wave spatial channels with a relatively small number of compressive measurements, in a manner that is compatible with coarse phase-only control and RF beamforming.  This allows scaling to a very large number of antenna elements without relying on channel reciprocity.  While our discussion of system design issues such as link budget and inter-cell beacon interference is tailored to outdoor 60 GHz picocellular networks, the basic approach is broadly applicable (e.g., to other bands, and to indoor environments). An important topic for future work is comprehensive experimental validation of our compressive approach,  which is based on a simple channel model including only the dominant rays.  At the network level, there are a host of design issues (e.g., see discussion in \cite{Zhu_Mobicom2014}).  The compressive approach allows each base station to build up an inventory of viable paths to nearby mobiles, but there is a huge design space to be explored on how base stations coordinate using this information to alleviate the effects of blockage (mobiles in urban environments can be routinely blocked by  pedestrians, automobiles, trees and other obstacles), and to manage inter- and intra-cell interference. Optimization of arrays of subarrays in base station ``faces'' for communicating with multiple users, as well as for handling channel dispersion, presents interesting design challenges.

\section*{Acknowledgement}
This work was supported by the National Science Foundation through the grant CNS-1317153, by the Institute for Collaborative Biotechnologies through the grant W911NF-09-0001 from the U.S. Army Research Office and by the Systems on Nanoscale Information fabriCs (SONIC), one of six centers supported by the STARnet phase of the Focus Center Research Program (FCRP), a Semiconductor Research Corporation program sponsored by MARCO and DARPA.  The content of the information does not necessarily reflect the position or the policy of the Government, and no official endorsement should be inferred

\bibliographystyle{IEEEtran}
\bibliography{references}

\begin{thebibliography}{10}
\providecommand{\url}[1]{#1}
\csname url@samestyle\endcsname
\providecommand{\newblock}{\relax}
\providecommand{\bibinfo}[2]{#2}
\providecommand{\BIBentrySTDinterwordspacing}{\spaceskip=0pt\relax}
\providecommand{\BIBentryALTinterwordstretchfactor}{4}
\providecommand{\BIBentryALTinterwordspacing}{\spaceskip=\fontdimen2\font plus
\BIBentryALTinterwordstretchfactor\fontdimen3\font minus
  \fontdimen4\font\relax}
\providecommand{\BIBforeignlanguage}[2]{{%
\expandafter\ifx\csname l@#1\endcsname\relax
\typeout{** WARNING: IEEEtran.bst: No hyphenation pattern has been}%
\typeout{** loaded for the language `#1'. Using the pattern for}%
\typeout{** the default language instead.}%
\else
\language=\csname l@#1\endcsname
\fi
#2}}
\providecommand{\BIBdecl}{\relax}
\BIBdecl

\bibitem{mobiledata}
\url{http://www.techjournal.org/2011/09/mobile-broadband-useage-is-set-to-explode-infographic}.

\bibitem{ramasamy_allerton12}
D.~Ramasamy, S.~Venkateswaran, and U.~Madhow, ``Compressive tracking with
  1000-element arrays: A framework for multi-{Gbps} mm wave cellular
  downlinks,'' in \emph{Communication, Control, and Computing (Allerton), 2012
  50th Annual Allerton Conference on}, Oct 2012.

\bibitem{Zhu_Mobicom2014}
Y.~Zhu, Z.~Zhang, Z.~Marzi, C.~Nelson, U.~Madhow, B.~Y. Zhao, and H.~Zheng,
  ``Demystifying 60ghz outdoor picocells,'' in \emph{Proceedings of the 20th
  Annual International Conference on Mobile Computing and Networking}, ser.
  MobiCom '14, 2014.

\bibitem{Valdes_2010}
A.~Valdes-Garcia, S.~Nicolson, J.-W. Lai, A.~Natarajan, P.-Y. Chen,
  S.~Reynolds, J.-H. Zhan, D.~Kam, D.~Liu, and B.~Floyd, ``A fully integrated
  16-element phased-array transmitter in sige {BiCMOS} for 60-ghz
  communications,'' \emph{Solid-State Circuits, IEEE Journal of}, vol.~45,
  no.~12, pp. 2757--2773, Dec 2010.

\bibitem{Cohen_2012}
E.~Cohen, M.~Ruberto, M.~Cohen, O.~Degani, S.~Ravid, and D.~Ritter, ``A {CMOS}
  bidirectional 32-element phased-array transceiver at 60ghz with ltcc
  antenna,'' in \emph{Radio Frequency Integrated Circuits Symposium (RFIC),
  2012 IEEE}, June 2012, pp. 439--442.

\bibitem{Rajagopal_Globecom_2011}
S.~Rajagopal, S.~Abu-Surra, Z.~Pi, and F.~Khan, ``Antenna array design for
  multi-{Gbps} mmwave mobile broadband communication,'' in \emph{Global
  Telecommunications Conference (GLOBECOM 2011), 2011 IEEE}, Dec 2011.

\bibitem{Heath_asilomar2012}
S.~Akoum, O.~El~Ayach, and R.~Heath, ``Coverage and capacity in mmwave cellular
  systems,'' in \emph{Signals, Systems and Computers (ASILOMAR), 2012
  Conference Record of the Forty Sixth Asilomar Conference on}, Nov 2012.

\bibitem{Rappaport_38G}
T.~Rappaport, Y.~Qiao, J.~Tamir, J.~Murdock, and E.~Ben-Dor, ``Cellular
  broadband millimeter wave propagation and angle of arrival for adaptive beam
  steering systems (invited paper),'' in \emph{Radio and Wireless Symposium
  (RWS), 2012 IEEE}, Jan 2012, pp. 151--154.

\bibitem{Rappaport_WCNC2012}
J.~Murdock, E.~Ben-Dor, Y.~Qiao, J.~Tamir, and T.~Rappaport, ``A 38 ghz
  cellular outage study for an urban outdoor campus environment,'' in
  \emph{Wireless Communications and Networking Conference (WCNC), 2012 IEEE},
  April 2012, pp. 3085--3090.

\bibitem{Rappaport_JSAC2014}
A.~Ghosh, T.~Thomas, M.~Cudak, R.~Ratasuk, P.~Moorut, F.~Vook, T.~Rappaport,
  G.~MacCartney, S.~Sun, and S.~Nie, ``Millimeter-wave enhanced local area
  systems: A high-data-rate approach for future wireless networks,''
  \emph{Selected Areas in Communications, IEEE Journal on}, June 2014.

\bibitem{Berraki_WCNC2014}
D.~Berraki, S.~Armour, and A.~Nix, ``Application of compressive sensing in
  sparse spatial channel recovery for beamforming in mmwave outdoor systems,''
  in \emph{Wireless Communications and Networking Conference (WCNC), 2014
  IEEE}, April 2014.

\bibitem{Junyi_JSAC2009}
J.~Wang, Z.~Lan, C.-W. Pyo, T.~Baykas, C.-S. Sum, M.~Rahman, J.~Gao, R.~Funada,
  F.~Kojima, H.~Harada, and S.~Kato, ``Beam codebook based beamforming protocol
  for multi-{Gbps} millimeter-wave {WPAN} systems,'' \emph{Selected Areas in
  Communications, IEEE Journal on}, vol.~27, no.~8, pp. 1390--1399, October
  2009.

\bibitem{david_love}
S.~Hur, T.~Kim, D.~J. Love, J.~V. Krogmeier, T.~A. Thomas, and A.~Ghosh,
  ``Millimeter wave beamforming for wireless backhaul and access in small cell
  networks,'' \emph{Communications, IEEE Transactions on}, vol.~61, no.~10, pp.
  4391--4403, 2013.

\bibitem{Heath_JSTSP2014}
A.~Alkhateeb, O.~El~Ayach, G.~Leus, and R.~Heath, ``Channel estimation and
  hybrid precoding for millimeter wave cellular systems,'' \emph{Selected
  Topics in Signal Processing, IEEE Journal of}, Oct 2014.

\bibitem{Heath_SPAWC_2014}
------, ``Single-sided adaptive estimation of multi-path millimeter wave
  channels,'' in \emph{Signal Processing Advances in Wireless Communications
  (SPAWC), 2014 IEEE 15th International Workshop on}, June 2014.

\bibitem{Tadilo_Globecom2014}
T.~Bogale and L.~B. Le, ``Beamforming for multiuser massive mimo systems:
  Digital versus hybrid analog-digital,'' in \emph{Global Communications
  Conference (GLOBECOM), 2014 IEEE}, Dec 2014, pp. 4066--4071.

\bibitem{torkildson}
C.~Sheldon, M.~Seo, E.~Torkildson, M.~Rodwell, and U.~Madhow, ``Four-channel
  spatial multiplexing over a millimeter-wave line-of-sight link,'' in
  \emph{Microwave Symposium Digest, 2009. MTT'09. IEEE MTT-S
  International}.\hskip 1em plus 0.5em minus 0.4em\relax IEEE, 2009, pp.
  389--392.

\bibitem{Gaojian2014}
G.~Wang and G.~Ascheid, ``Joint pre/post-processing design for large millimeter
  wave hybrid spatial processing systems,'' in \emph{European Wireless 2014;
  20th European Wireless Conference; Proceedings of}, May 2014.

\bibitem{Singh_Globecom2014}
J.~Singh and S.~Ramakrishna, ``On the feasibility of beamforming in millimeter
  wave communication systems with multiple antenna arrays,'' in \emph{Global
  Communications Conference (GLOBECOM), 2014 IEEE}, Dec 2014.

\bibitem{Heath_ICC2012}
O.~Ayach, R.~Heath, S.~Abu-Surra, S.~Rajagopal, and Z.~Pi, ``Low complexity
  precoding for large millimeter wave mimo systems,'' in \emph{Communications
  (ICC), 2012 IEEE International Conference on}, June 2012.

\bibitem{Heath_SPAWC2012}
O.~El~Ayach, R.~Heath, S.~Abu-Surra, S.~Rajagopal, and Z.~Pi, ``The capacity
  optimality of beam steering in large millimeter wave mimo systems,'' in
  \emph{Signal Processing Advances in Wireless Communications (SPAWC), 2012
  IEEE 13th International Workshop on}, June 2012.

\bibitem{zhang2010statistical}
H.~Zhang and U.~Madhow, ``Statistical modeling of fading and diversity for
  outdoor 60 ghz channels,'' in \emph{Proceedings of the 2010 ACM international
  workshop on mmWave communications: from circuits to networks}.

\bibitem{zhang2010channel}
H.~Zhang, S.~Venkateswaran, and U.~Madhow, ``Channel modeling and mimo capacity
  for outdoor millimeter wave links,'' in \emph{Wireless Communications and
  Networking Conference (WCNC), 2010 IEEE}.

\bibitem{ramasamy_ita12}
D.~Ramasamy, S.~Venkateswaran, and U.~Madhow, ``Compressive adaptation of large
  steerable arrays,'' in \emph{Information Theory and Applications Workshop
  (ITA), 2012}, feb. 2012.

\bibitem{ramasamy_asilomar12}
------, ``Compressive estimation in {AWGN}: General observations and a case
  study,'' in \emph{Signals, Systems and Computers (ASILOMAR), 2012 Conference
  Record of the Forty Sixth Asilomar Conference on}, Nov 2012.

\bibitem{Ramasamy_TSP_2014}
------, ``Compressive parameter estimation in {AWGN},'' \emph{Signal
  Processing, IEEE Transactions on}, April 2014.

\bibitem{achlioptas_database-friendly_2001}
D.~Achlioptas, ``Database-friendly random projections,'' ser. {PODS} '01.\hskip
  1em plus 0.5em minus 0.4em\relax New York, {NY}, {USA}: {ACM}, 2001.

\bibitem{baraniuk_simple_RIP_2008}
R.~Baraniuk, M.~Davenport, R.~{DeVore}, and M.~Wakin, ``A simple proof of the
  restricted isometry property for random matrices,'' \emph{Constructive
  Approximation}, 2008.

\bibitem{VanTrees:Bell:Book:Bounds}
H.~L.~V. Trees and K.~L. Bell, \emph{Bayesian Bounds for Parameter Estimation
  and Nonlinear Filtering/Tracking}, 2007.

\bibitem{Basu::ZZB::Periodic::2000}
S.~Basu and Y.~Bresler, ``A global lower bound on parameter estimation error
  with periodic distortion functions,'' \emph{Information Theory, IEEE
  Transactions on}, 2000.

\end{thebibliography}

\end{document}